\documentclass[aip,graphicx]{revtex4-1}

\usepackage{comment, color, graphicx}

\newcommand{\vv}{v}
\newcommand{\vpar}{\vv_{||}}
\newcommand{\Vpar}{V_{||}}
\newcommand{\ang}{\chi}
\newcommand{\aang}{\alpha}
\newcommand{\vect}[1]{\mbox{\boldmath $#1$}}
\newcommand{\iotab}{{\rlap{-} {\iota }}}

\newcommand{\sgn}{\mathrm{sgn}}

\newcommand{\be}{\begin{equation}}
\newcommand{\ee}{\end{equation}}

\newcommand{\p}{\partial}

\newcommand{\zmax}{\hat{\zeta}}
\newcommand{\thmax}{\hat{\theta}}
\newcommand{\Bmax}{\hat{B}}
\newcommand{\Bmin}{\check{B}}

\newcommand{\vm}{\vect{v}_m}
\newcommand{\iotah}{\tilde{\iota}}
\newcommand{\zetah}{\tilde{\zeta}}
\newcommand{\thetah}{\tilde{\theta}}
\newcommand{\Np}{N_p}
\newcommand{\PS}{Pfirsch-Schl\"{u}ter~}
\newcommand{\W}{W}
\newcommand{\D}{D}
\newcommand{\vperp}{v_\bot}

\newcommand{\Er}{E_r}
\newcommand{\gCS}{\mathcal{S}}
\newcommand{\F}{\mathcal{F}}
\newcommand{\G}{\mathcal{G}}
\newcommand{\Gbs}{{\langle G_{bs} \rangle}}
\newcommand{\y}{{\mathcal{Y}}}
\newcommand{\Dz}{{\Delta_\zeta}}
\newcommand{\ptot}{p_{\Sigma}}
\draft 

\begin{document}


\title{Omnigenity as generalized quasisymmetry}



\author{Matt Landreman}
\email[]{landrema@mit.edu}
\author{Peter J. Catto}
\affiliation{Plasma Science and Fusion Center, MIT, Cambridge, MA, 02139, USA}


\date{\today}

\begin{abstract}
Any viable stellarator reactor will need to be nearly omnigenous, meaning the radial guiding-center drift
velocity averages to zero over time for all particles.  While omnigenity is easier to achieve than
quasisymmetry, we show here that several properties of quasisymmetric plasmas also apply directly or with only minor modification to the larger
class of omnigenous plasmas.  For example, concise expressions exist
for the flow and current, closely resembling those for a tokamak, and these expressions are explicit in that no magnetic differential equations
remain.  A helicity $(M,N)$ can be defined for any omnigenous field, based on the topology by which $B$ contours close on
a flux surface, generalizing the helicity
associated with quasisymmetric fields.  For generalized quasi-poloidal symmetry ($M=0$),
the bootstrap current vanishes, which may yield desirable equilibrium and stability properties.
A concise expression
is derived for the radial electric field in any omnigenous plasma that is not quasisymmetric.
The fact that tokamak-like analytical calculations are possible in omnigenous plasmas despite their fully-3D magnetic spectrum
makes these configurations useful for gaining insight and benchmarking codes.
A construction is given to produce omnigenous $B(\theta,\zeta)$ patterns with stellarator symmetry.

\end{abstract}


\maketitle 


\section{Introduction}

Nonaxisymmetric toroidal magnetic fields can provide intrinsically steady-state, disruption-free plasma confinement.
However, unlike axisymmetric fields, nonaxisymmetric fields do not generally confine all trapped particle orbits.  This shortcoming has a modest deleterious
effect on energy confinement, which still follows scalings comparable to tokamak confinement due to turbulent transport \cite{ISS04}. However, unconfined orbits
will likely pose a serious problem in a reactor, where unconfined energetic alpha particles will collide with the first wall before thermalizing,
causing unacceptable damage to the plasma-facing components.
Consequently, one criterion of stellarator optimization in recent years has been the quality of collisionless particle confinement.
The ideal limit in which all collisionless trajectories are confined is known as omnigenity (sometimes spelled omnigeneity).
Omnigenity can be defined more precisely as the condition that the time average of $\vm\cdot\nabla\psi$ along each field line vanishes for
all values of magnetic moment $\mu=v_\bot ^2/(2B)$.  Here, $\vm = \vpar^2 \vect{B}\times\vect{\kappa}/(\Omega B) + \vperp^2 (\vect{B}\times\nabla B)/(2\Omega B^2)$ is the magnetic drift velocity, $\vect{\kappa}=\vect{b}\cdot\nabla\vect{b}$, $\vect{b}=\vect{B}/B$, $B=|\vect{B}|$, $\Omega = ZeB/(mc)$ is the gyrofrequency,
$Z$ is the species charge in units of the proton charge $e$, $m$ is the mass, $c$ is the speed of light, and $2\pi\psi$ is the toroidal flux.  As shown in Appendix \ref{a:proofs},
an equivalent definition of omnigenity is that the longitudinal adiabatic invariant $J=\oint \vpar d\ell$ is a constant on a flux surface.
Other equivalent definitions are the absence of a ``$1/\nu$" regime of confinement, where $\nu$ is the collision frequency, or
an effective helical ripple of zero.

One method of obtaining omnigenity in a nonaxisymmetric toroidal system is quasisymmetry,
the design principle behind the HSX\cite{HSX} and NCSX\cite{NCSX} experiments.
Quasisymmetry is usually defined\cite{Boozer95} as the condition that the field magnitude $B$ varies on a flux surface only
through a fixed linear combination of the poloidal and toroidal Boozer angles:
\begin{equation}
\label{eq1}
B=B(\psi, \, M\theta-N\zeta)
\end{equation}
 for integers $M$ and $N$.
Quasisymmetry can also be defined as
$B=B(\psi, \, M\theta_*-N\zeta_*)$ for other coordinates $(\theta_*, \, \zeta_*)$ such as Hamada angles
(as proved in Appendix \ref{a:Hamada}), or by the coordinate-free conditions\cite{PerAndreiPRL}
 $\vect{B}\cdot\nabla \left[ (\vect{B}\times\nabla\psi\cdot\nabla B) / (\vect{B}\cdot\nabla B) \right]=0$
or\cite{AndreiPer2011} $\nabla B \times\nabla \psi \cdot\nabla(\vect{B}\cdot\nabla B)=0$ .
The Lagrangian $L$ for guiding-center drift motion, when expressed in Boozer coordinates\cite{CSPoP},
 only depends on $B$ and not the vector components of $\vect{B}$. Consequently, an ignorable coordinate in $B$ gives rise to a conservation law.
 The conserved quantity\cite{Boozer83, Zille}, resembling canonical angular momentum, ensures each particle can only drift a distance on the order of
 a gyroradius away from a given flux surface, implying omnigenity.

While every quasisymmetric field is omnigenous, not every omnigenous field is quasisymmetric.
A proof by Cary and Shasharina at first appears to show the opposite conclusion,
that a field which is both infinitely differentiable (analytic) and perfectly omnigenous must be quasisymmetric \cite{CSPoP, CSPRL}.
However, the same authors point out that the proof is quite fragile: an analytic field that deviates from omnigenity only slightly
may depart from quasisymmetry by a great amount.
(We will construct examples of such fields in Section \ref{sec:construction}.) Thus, in practice, omnigenity does
not imply quasisymmetry.  Consequently, to achieve good collisionless particle confinement, there is no need to strive for the
strict condition of quasisymmetry when the weaker condition of omnigenity is sufficient.  While it is possible to
find three-dimensional MHD equilibria that are reasonably close to quasisymmetry\cite{Zille, HSX},
lifting the demand of quasisymmetry widens the parameter space, allowing stellarators to be better optimized for other criteria.

One may still aim to achieve quasisymmetry for a different reason:
In non-quasisymmetric plasmas, the parallel flow is determined by leading-order neoclassical processes,
while in quasisymmetric plasmas, the parallel flow is determined by turbulence\cite{PerAndreiPRL}.
As flows and flow shear may affect MHD modes and microinstabilities,
quasisymmetric plasmas may have unique stability and turbulence properties,
though more research is required to explore this notion.

Quasisymmetric fields provide an important and useful ideal limit for understanding nonaxisymmetric plasmas.
In quasisymmetric fields, concise analytic expressions can be derived for the neoclassical distribution function,
radial fluxes, and parallel flows and current \cite{Pytte, Boozer83}.
These formulae are nearly identical to the corresponding formulae for a tokamak.  The formulae are
explicit, in that they do not involve solutions of partial differential equations.
In contrast, the corresponding formulae for a general stellarator must be expressed in terms of the
solutions of partial differential equations involving the field strength \cite{ShaingCallen, BoozerBootstrap, Wakatani, Johnson}.
One example is the system of equations (\ref{eqb2})-(\ref{eqb5}) for the bootstrap current.

It is the purpose of this paper to demonstrate that perfect omnigenity is another useful ideal limit.
Omnigenity, while less restrictive than quasisymmetry, is still a strong enough condition
to place powerful constraints on $B(\theta,\zeta)$. For example \cite{CSPoP},
in the neighborhood of each minimum and maximum of $B$ along a field line,
each adjacent field line segment on the flux surface has an extremum at the same value of $B$,
as we will show in Section \ref{sec:BProperties} using novel arguments.
We will also show that field lines can never
be parallel to the contours of $B$ on each flux surface,
and these contours must topologically
link the flux surface toroidally, poloidally, or both.
By defining integers $M$ and $N$ as the number of times each contour links the plasma toroidally and poloidally respectively,
the helical ``mode numbers" $M$ and $N$ in (\ref{eq1})
can be generalized to any omnigenous plasma.  Using this generalization,
we will show in Section \ref{sec:parallelTransport} that formulae for the bootstrap current and flux-surface-averaged parallel flow
in a quasisymmetric plasma in fact apply to the larger class
of \emph{all} omnigenous plasmas.  The formulae for the \PS flow and current in a
quasisymmetric plasma also apply to all omnigenous plasmas if a new term is added,
consisting of an integral of derivatives of $B$.
The same is true of the low-collisionality distribution functions.
Some of these neoclassical properties were discussed previously in Refs. \onlinecite{Subbotin}, \onlinecite{hn}, and \onlinecite{meQI}
for the specific case of generalized poloidal symmetry ($M=0$).  Here we give
more general calculations that apply to omnigenity of any helicity.
In Refs. \onlinecite{Subbotin} and \onlinecite{hn} it was pointed out that for an $M=0$ omnigenous plasma,
the bootstrap current vanishes unless there is inductive or RF-driven current.  This result will be recovered from the
analysis of general-helicity omnigenous fields here.

Another interesting feature of omnigenous magnetic fields that we will demonstrate concerns the radial electric field $\Er$.
In a general stellarator, the radial electron and ion particle fluxes exhibit different dependencies on $\Er$.
Consequently, it is possible to determine the electric field using the condition that the net radial
current must vanish, i.e. quasineutrality or ambipolarity.
However, this procedure fails in tokamaks or in quasisymmetric stellarators, which possess
the property of ``intrinsic ambipolarity."  This is the property that to leading order
in the expansion of gyroradius to system size, the
net radial current vanishes regardless of the radial electric field, and so the
electric field is determined by higher-order processes that are more difficult to calculate.
An omnigenous plasma that is far from quasisymmetry will not be intrinsically ambipolar, and so it is still possible
to solve for $\Er$ using ambipolarity.  In Section \ref{sec:Er} we will derive $\Er$ for any omnigenous, non-quasisymmetric field.
The result differs from previously known expressions for the electric field in non-omnigenous stellarators.
Our calculation will also show from a new perspective that $\Er$
indeed becomes undetermined in the limit of quasisymmetry.

Finally, in section \ref{sec:construction}, we will derive some further geometric properties of $B(\theta,\zeta)$ for an omnigenous
flux surface.  It will be shown how to generate a family of omnigenous flux surfaces that are consistent with an additional symmetry usually
possessed by stellarator experiments.

\section{Magnetic field properties}
\label{sec:BProperties}

\subsection{Extrema of $B$ on each flux surface}
\label{sec:const}

Let us now begin to analyze the geometric properties of omnigenous fields in detail.
Let $\Bmin(\vect{r})$ and $\Bmax(\vect{r})$ denote the nearest minimum and maximum of $B$ found by moving forward and backward along a field line from
the starting position $\vect{r}$.  In a general stellarator, $\Bmin$ and $\Bmax$ will take on a continua of values (or several continua) on each flux surface.
However, in an omnigenous field we will now show that $\Bmin$ and $\Bmax$ may only take on \emph{discrete} values on each flux surface, and
in the simplest case, each has only a single value on each flux surface.
In other words, in the neighborhood of an extremum of $B$ along a field line,
each nearby field line segment on the same flux surface has an extremum at the same value  of $B$.

Let us first prove this property for the minimum $\Bmin$.
We begin by recalling the contravariant and covariant expressions\cite{Boozer81} for $B$ in terms of the poloidal Boozer angle $\theta$ and the toroidal
Boozer angle $\zeta$:
\begin{equation}
\vect{B} = \nabla \psi \times \nabla \theta + \iotab \nabla \zeta \times \nabla \psi,
\label{contravariant}
\end{equation}
\begin{equation}
\vect{B} = \beta(\psi, \theta, \zeta) \nabla\psi + I(\psi) \nabla \theta + G(\psi) \nabla \zeta.
\label{covariant}
\end{equation}
Here, $\iotab$ is the rotational transform, $I(\psi)$ is $2/c$ times the toroidal current inside the flux surface $\psi$,
and $G(\psi)$ is $2/c$ times the poloidal current outside the flux surface $\psi$.
Now let $\vect{r}_0 = (\theta_0, \zeta_0)$ denote a point at which $B$ is minimized with respect to movement along a field line.
At this point,
$\vect{B}\cdot\nabla B$ must vanish, so
\begin{equation}
0=(\vect{B}\cdot\nabla\theta) (\partial B/\partial\theta) + (\vect{B}\cdot\nabla\zeta) (\partial B/\partial\zeta).
\label{eq:par0}
\end{equation}
If the field is omnigenous, deeply trapped particles at $\vect{r}_0$ must have no radial drift.  As
 $\vm \cdot\nabla\psi \propto \vect{B}\times\nabla B\cdot\nabla\psi$,
then
\begin{equation}
0=\vect{B}\times\nabla B \cdot \nabla\psi = - G (\vect{B}\cdot\nabla\zeta) (\partial B/\partial\theta) + I (\vect{B}\cdot\nabla\zeta )(\partial B/\partial\zeta).
\label{eq:drift0}
\end{equation}
Equations (\ref{eq:par0}) and (\ref{eq:drift0}) are a system of two linearly independent equations
for $\p B/\p \theta$ and $\p B/\p \zeta$ at $\vect{r}_0$,
so both of these derivatives must be zero.
The vanishing of these derivatives implies $B(\vect{r}_0)$ is either an isolated local minimum, a saddle point, or one point along
a ``valley" of constant $\Bmin$.

The first two of these three possibilities can be excluded as follows. We first Taylor-expand
\begin{equation}
\label{eq:taylor}
B \approx B_0 + \frac{x}{2}(\theta-\theta_0)^2
+y(\theta-\theta_0)(\zeta-\zeta_0)
+ \frac{z}{2}(\zeta-\zeta_0)^2,
\end{equation}
where
$x=\left[\p^2 B/\p \theta^2\right]_0$,
$y=\left[\p^2 B/\p \theta\, \p\zeta\right]_0$,
$z=\left[\p^2 B/\p \zeta^2\right]_0$,
and the zero subscripts indicate quantities evaluated at $\vect{r}_0$.
Considering the variation of $B$ along the nearby field line $\theta = \theta_0 + \iotab (\zeta-\zeta_0) + \delta$ for some small $\delta$,
by eliminating either $\theta$ or $\zeta$ in (\ref{eq:taylor}) and completing the square, it is evident that $B$ is minimized along the
field line at the point $\vect{r}_1=(\theta_1, \zeta_1)$, where
$\theta_1 = \theta_0 +(\iotab y+z)\delta/A $,
$\zeta_1 = \zeta_0 -(\iotab x + y)\delta/A$,
and $A=\iotab^2 x+2\iotab y+z$.
Plugging the definitions of $\theta_1$ and $\zeta_1$ into (\ref{eq:taylor}) gives
\begin{equation}
B \approx B_0 + \frac{k\delta^2}{2A}+(\theta-\theta_1)\frac{k\delta}{A} - (\zeta-\zeta_1)\frac{\iotab k \delta}{A}
+\frac{x}{2}(\theta-\theta_1)^2
+y(\theta-\theta_1)(\zeta-\zeta_1)
+ \frac{z}{2}(\zeta-\zeta_1)^2
\label{eq:shiftedTaylor}
\end{equation}
where $k=xz-y^2$.
Since $B$ is minimized along the shifted field line at $\vect{r}_1$, the radial drift must vanish there,
for the same reason it had to vanish at $\vect{r}_0$, so
$\p B/\p \theta=0$ and $\p B/\p \zeta=0$ at $\vect{r}_1$.  Consequently the third and fourth right-hand-side terms in
(\ref{eq:shiftedTaylor}) (those
linear in $(\theta-\theta_1)$ and $(\zeta-\zeta_1)$)  must vanish, and so $k$ must be zero. Thus, the second term on the right hand side of
(\ref{eq:shiftedTaylor}) vanishes, and so the minimum of $B$ along the shifted field line is $B_0$,
the same as the minimum on the original field line.  This proves the desired result for $\Bmin$.
This fact, that deeply trapped particles are perfectly confined when all $\Bmin$ on a flux surface are the same,
was first observed in Ref. \onlinecite{MynickChuBoozer}.

Now we make the analogous argument for $\Bmax$.  Let $\vect{r}_0$ now represent a point
at which $B$ is maximized along a field line.  The bounce time diverges for barely trapped particles, because they spend an infinitely
long time near the turning points. Thus, if there is an outward radial drift at $\vect{r}_0$, marginally trapped particles
there will have an arbitrarily large radial excursion.
Even if these particles would in principle make a large inward radial step
at the opposite bounce point, they would drift out of the machine before having time to get
to the other bounce point, so these particles effectively would have a nonzero time-averaged radial drift.
We choose to include in the definition of omnigenity the condition that marginally trapped
particles cannot have radial steps of unbounded size in this manner.
Now suppose at $\vect{r}_0$ there were an \emph{inward} radial drift.
By omnigenity, marginally trapped particles must make an
arbitrarily large outward excursion elsewhere in the trajectory to balance the arbitrarily large inward
excursion in the neighborhood of $\vect{r}_0$, so this case too is unacceptable.
Therefore, $\vm\cdot\nabla\psi$ must vanish at $\vect{r}_0$,
implying (\ref{eq:drift0}).  The rest of the argument for the constancy of $\Bmin$ then applies, and so $\Bmax$ must be constant as well for each field
line in a neighborhood of $\vect{r}_0$ on the flux surface.

\begin{figure}
\includegraphics{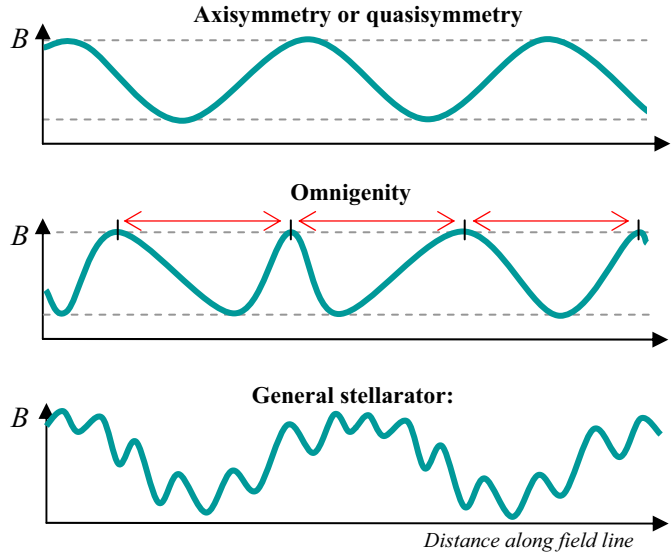}
\caption{(Color online) Omnigenous fields have an intermediate level of complexity between
quasisymmetric and general nonaxisymmetric fields.
\label{fig:BAlongFieldLine}}
\end{figure}

Due to these constraints on the extrema of $B$, omnigenous fields represent
an intermediate level of complexity between quasisymmetric fields and general nonaxisymmetric fields.
Figure \ref{fig:BAlongFieldLine} illustrates this point. For axisymmetric and quasisymmetric fields, the $B$ wells in which particle are trapped
all have identical shape.  At the opposite extreme of a general three-dimensional field,
different field line segments have different maxima and minima of $B$.
Omnigenous fields represent a middle ground.  Maxima of $B$ occur repeatedly at the same values of $B$, and the same is true of the minima.
In fact, we will prove in Section \ref{sec:BMaxIsStraight} that the maxima are equally spaced (in $\theta$, in $\zeta$, and in distance along the field),
as indicated by the horizontal red arrows in Figure \ref{fig:BAlongFieldLine}.
Yet, unlike in quasisymmetric fields, the shape of the $B$ wells is different at different points along a field line.

\subsection{Generalized helicity}
\label{sec:helicity}

\begin{figure}
\includegraphics{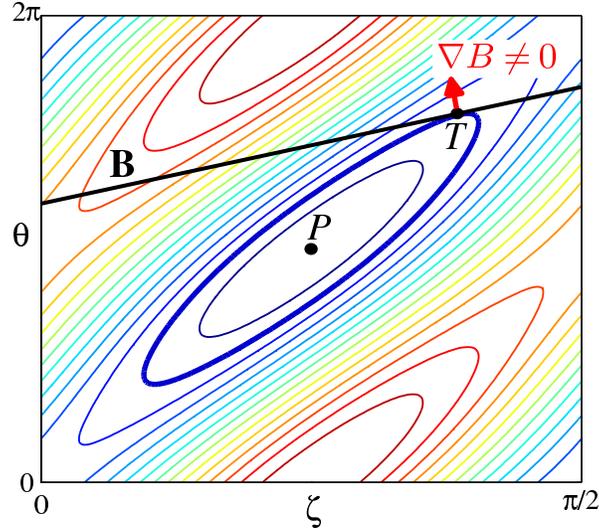}
\caption{(Color online) Contours of $B$ on a flux surface of a non-omnigenous stellarator. Contours typically
exist that do not topologically link the flux surface, such as the bold contour here.
The bold black straight line is a field line.
\label{fig:nonOmni}}
\end{figure}

\begin{figure}
\includegraphics{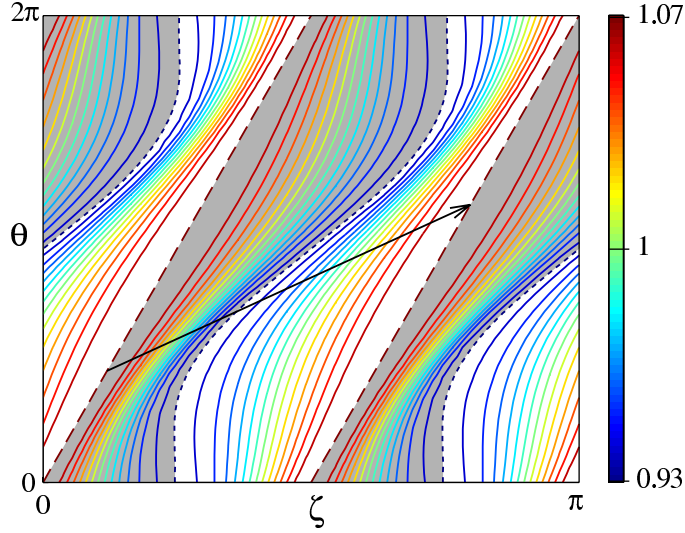}
\caption{(Color online) The field magnitude $B/\langle B^2 \rangle ^{1/2}$ for two of the four periods of an omnigenous field
with $M=1$, $N=4$, and $\iotab = 1.05$.
The straight dashed line is the maximum and the curved dotted curve is the minimum.
Field lines are parallel to the arrow.  One branch is shaded and the other is unshaded.
\label{fig:B}}
\end{figure}

\begin{figure}
\includegraphics{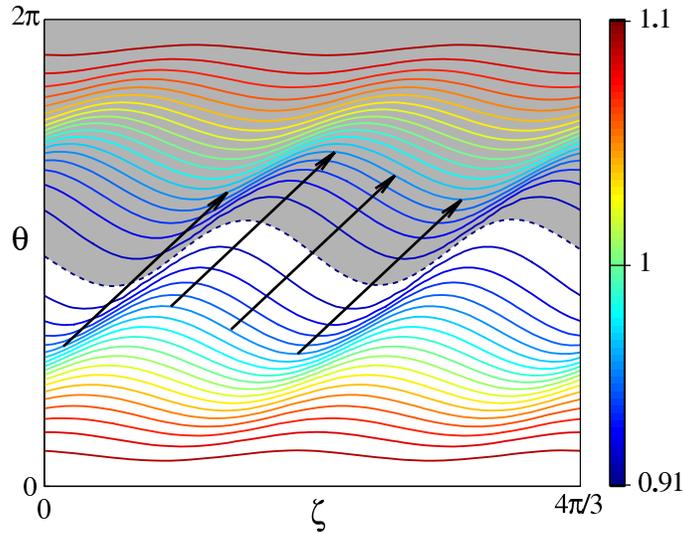}
\caption{(Color online) The field magnitude $B/\langle B^2 \rangle ^{1/2}$ for two of the three periods of an omnigenous field
with $M=1$, $N=0$ (generalized quasi-axisymmetry), and $\iotab = 1.62$.  Markings are as defined in Figure \ref{fig:B}.
The arrows point along $\vect{B}$ and connect two contours of equal $B$.
The conditions $\p \Delta_\theta/\p\ang=0$ and $\p\Delta_\zeta /\p\ang=0$ imply the arrows must
have the same length.
\label{fig:B2}}
\end{figure}

In an omnigenous toroidal field, the constant-$B$ contours on a flux surface must all encircle
the plasma toroidally, poloidally, or both.  In other words, each contour must topologically link
the flux surface: a contour cannot be continuously deformable (homotopic) to a point without leaving the flux surface.
If any constant-$B$ contour did not link the flux surface in this manner,
then the contour would enclose a \emph{point} maximum or minimum of $B$ within the surface,
and we just proved that such point extrema cannot occur in an omnigenous field.
As an illustration, the bold curve in Figure \ref{fig:nonOmni} is
a $B$ contour that does not link the plasma.  This contour
encloses a point minimum or maximum of $B(\theta,\zeta)$ at $P$,
and such points cannot exist in an omnigenous field.
For contrast, Figure \ref{fig:B} depicts an omnigenous field, one in which the $B$ contours
encircle the plasma both poloidally and toroidally.
Figure \ref{fig:B2} shows an omnigenous field with toroidally closed $B$ contours.
Neither field has any point extrema of $B$.

We can also prove that all $B$ contours must encircle the plasma using the following alternative argument.
If a constant-$B$ contour does not link the flux surface, then
there will be points at which the contour is tangent to the field, such
as point $T$ in Figure \ref{fig:nonOmni}. (Recall that field lines are straight
in the $(\theta,\zeta)$ plane for Boozer coordinates.)  At such a point of tengency,
$\vect{B}\cdot\nabla B=0$, but the derivative of $B$ in any other direction on  the flux surface
is nonzero. Thus, (\ref{eq:par0}) is satisfied while (\ref{eq:drift0}) is not,
violating the condition of omnigenity.  Physically,
if $B$ is a minimum along the field line at $T$,
deeply trapped particles at $T$ will have nonzero average radial drift,
while if $B$ is a maximum along the field line at $T$,
barely trapped particles will make an infinite radial step when they bounce at $T$.
Neither type of radial motion is
allowed in an omnigenous field, so the $B$ contours of each
flux surface in an omnigenous field can never be tangent to field lines.
This implies the contours must link the plasma.
It can be seen in Figures \ref{fig:B} and \ref{fig:B2} that the $B$ contours
are indeed nowhere tangent to the field lines.

We can now define integers $M$ and $N$ as follows:
each $B$ contour closes on itself after traversing the torus $M$ times toroidally and $N$ times poloidally, that is, after $\theta $
increases by $2\pi N$ and $\zeta $ increases by $2\pi M$.  This convention may seem backward at first,
but it is consistent with the $M$ and $N$ defined earlier for a quasisymmetric field: $B=B(M\theta-N\zeta)$.
By defining $M$ and $N$ in terms of the topology of the $B$ contours as we have done,
the helicity associated with quasisymmetric fields is generalized to any omnigenous field.
Notice that this helicity is completely independent of the rotational transform $\iotab$.

It should be noted that definition of the term ``quasi-isodynamic" given by some authors\cite{hn}
is equivalent to the condition ``omnigenous with $M=0$."
However, other authors\cite{MynickReview} define ``quasi-isodynamic" differently,
as the case in which only particles with a particular value of normalized magnetic moment
$\lambda=\vv_\bot^2/(\vv^2 B)$ are omnigenous, not all particles.

Next, it will be convenient to define a field line label $\ang$ by
\begin{equation}
\label{eq:ang}
\chi =(\theta -\iotab \zeta ) / (N-\iotab M).
\end{equation}
This definition is convenient because if a constant-$B$ curve is followed until it closes on itself,
then $\ang$ will increase by $2\pi $.  For much of the analysis that follows we will use $\left( {\psi ,\chi ,B} \right)$ coordinates.
In these coordinates the magnetic field has a contravariant form
$\vect{B}=\left( {N-\iotab M} \right)\nabla \psi \times \nabla \chi$
and a covariant form
\begin{equation}
\label{eq4}
\vect{B}=B_\psi \nabla \psi +B_B \nabla B+B_\chi \nabla \chi .
\end{equation}
The Jacobian is $\nabla\psi\times\nabla\ang\cdot\nabla B = (\vect{B}\cdot\nabla B)/(N-\iotab M)$.
We now derive several properties of the covariant coefficients.
The inner product of (\ref{eq4}) with ${\rm {\bf B}}$ gives
$B_B = B^2 / \vect{B}\cdot \nabla B$.
The inner product of (\ref{eq4}) with $\nabla \psi \times \nabla B$ gives
\begin{equation}
\label{eq6}
B_\chi =-\left( {N-\iotab M} \right)\frac{{\rm {\bf B}}\times \nabla \psi
\cdot \nabla B}{{\rm {\bf B}}\cdot \nabla B}.
\end{equation}
As discussed in Section \ref{sec:const}, the numerator of (\ref{eq6}) vanishes
whenever the denominator does, leaving $B_\ang$ nonsingular.  Indeed, since omnigenity
precludes $B$ contours from being tangent to field lines, then $\p \vect{r} / \p\ang$ is
never singular, and so $B_\ang = \vect{B} \cdot \p \vect{r} / \p\ang$ cannot be singular.
Here and throughout this paper, $\p /\p\ang$ is performed at fixed $B$, $\p/\p B$ is performed at
fixed $\ang$, $\p/\p\theta$ is performed at fixed $\zeta$, and $\p/\p\zeta$ is performed at fixed
$\theta$, unless denoted explicitly with a subscript.

\begin{figure}
\includegraphics[width=0.6\textwidth]{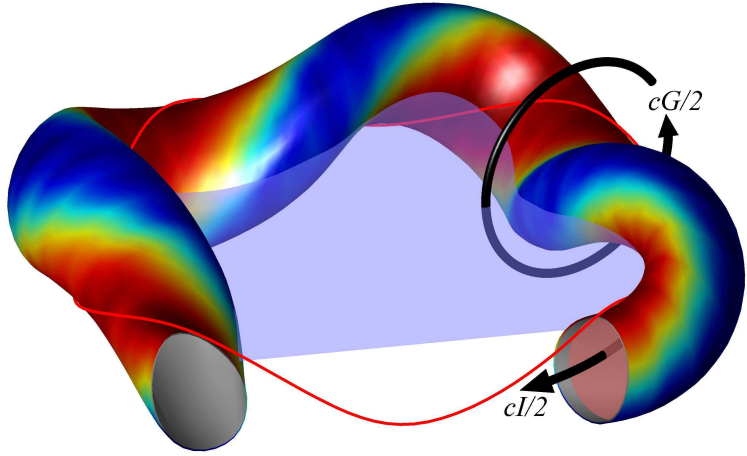}
\caption{(Color online) Equation (\ref{eq13}) is derived by applying Amp\`{e}re's Law to a $B$ contour,
such as the red curve here.  The black arrows illustrate the toroidal and poloidal currents,
which are the currents through the red and blue translucent surfaces respectively.
\label{fig:Ampere}}
\end{figure}

Now consider a path on a flux surface that follows a constant-$B$ curve
until it closes on itself. As described above, $\chi $ increases by $2\pi $
along this curve. From Amp\`{e}re's Law,
$(4\pi/c)i_h =\int {{\rm {\bf B}}\cdot d{\rm {\bf r}}} =\int_0^{2\pi }
{d\chi } \,{\rm {\bf B}}\cdot \partial {\rm {\bf r}} /\partial \chi
=\int_0^{2\pi } {B_\chi d\chi } $
where $i_h $ is the current linked by the loop, and the integrals are
performed at constant $\psi $ and $B$.
The integration path links the torus $N$ times poloidally and $M$ times poloidally, so
the amount of current linked by this helical curve is $i_h =Ni_t +Mi_p $
where $i_t $ is the toroidal current inside the flux surface and $i_p $ is
the poloidal current outside the flux surface. These currents are related to
the coefficients of the Boozer covariant representation
by $I=2i_t /c$ and $G=2i_p /c$. Therefore $i_h =c\left( MG+NI \right)/2$
and
$\int_0^{2\pi } {B_\chi d\chi } =2\pi \left( MG+NI \right)$.  This application of
Amp\`{e}re's Law is illustrated in Figure \ref{fig:Ampere} (for $M=1$ and $N=4$.)
Any single-valued function of position must be periodic in $\chi $ (with
period $2\pi )$ if $B$ is held fixed. In particular, $B_\chi $ must be
periodic in this way, so we can write
$B_\chi =MG+NI+ \partial h /\partial \chi$
for some single-valued $h$. Hence, recalling (\ref{eq6}), we obtain the useful formula
\begin{equation}
\label{eq13}
\frac{{\rm {\bf B}}\times \nabla \psi \cdot \nabla B}{{\rm {\bf B}}\cdot
\nabla B}=-\frac{1}{\left( {N-\iotab M} \right)}\left( {MG+NI+\frac{\partial
h}{\partial \chi }} \right).
\end{equation}

In the quasisymmetric limit $B\left( {\theta ,\zeta } \right)=B\left(
{M\theta -N\zeta } \right)$, then the left-hand side of this expression can be computed directly,
and the result is the same but without the $h$ term. Thus,
quasisymmetry corresponds to the $\p h/\p\ang=0$ limit.

\subsection{Relation between branches}
\label{sec:BMaxIsStraight}

The continuous coordinates $(\psi,\chi,B)$ do not uniquely determine a point in space,
both because there may be multiple toroidal segments, but also because within each segment there are two points
at given $B$ on either side of $\Bmin$, the minimum of $B$ on the flux surface. This discrete degree of freedom, called the ``branch," will be denoted by $\gamma=\pm 1$.
The shaded and unshaded regions of Figures \ref{fig:B} and \ref{fig:B2} illustrate the two branches.
The variations of $B$ in the two branches are related due to the condition of omnigenity. As shown in Appendix \ref{a:proofs},
\begin{equation}
\label{eq15}
\frac{\partial }{\partial \chi }\sum\limits_\gamma {\frac{\gamma }{{\rm {\bf
b}}\cdot \nabla B}} =0.
\end{equation}
This result was derived using different notation in Ref. \onlinecite{CSPoP} and is termed the ``Cary-Shasharina Theorem" in Ref. \onlinecite{hn}.

Using (\ref{eq15}), we can prove several facts about pairs of points on a same field line that share the same $B$ but lie on opposite sides of $\Bmin$.
Let $\Delta_\ell$ be the distance between these points, measured along the field line.
In a general stellarator, $\Delta_\ell$ will depend on the field line label $\ang$ (in addition to $B$ and $\psi$).
But since $\Delta_\ell = \sum_\gamma \gamma \int_{\Bmin}^B [(\vect{b}\cdot\nabla B)']^{-1} dB'$ (i.e. $\vect{b}\cdot\nabla B$ is evaluated
at $B'$ rather than $B$),
then (\ref{eq15}) implies $\p \Delta_\ell / \p\ang =0$ in an omnigenous field.  Thus, for any given $B$, every such pair of points
 has the same separation $\Delta_\ell$.
This result is illustrated for the case of $\Bmax$, the maximum of $B$ on the flux surface, by the horizontal red arrows in Figure \ref{fig:BAlongFieldLine}.
A similar result
holds for the difference in $\zeta$ between pairs of points.  Let $\Delta_\zeta = \sum_\gamma \gamma \int_{\Bmin}^B (\p\zeta/\p B)' dB'$ be the
difference in $\zeta$ between a pair of points as described above.  Notice $\p\zeta/\p B = (\vect{B}\cdot\nabla\zeta) / \vect{B}\cdot\nabla B$.
Multiplying the covariant and contravariant Boozer representations (\ref{contravariant})-(\ref{covariant}),
\begin{equation}
\label{eq20}
B^2 / (G+\iotab I) =  \nabla \psi \times \nabla \theta \cdot \nabla \zeta
={\rm {\bf B}}\cdot \nabla \zeta =\iotab ^{-1}{\rm {\bf B}}\cdot \nabla\theta.
\end{equation}
Therefore,
\begin{equation}
\label{eq20b}
\Delta_\zeta = \frac{1}{G+\iotab I} \sum_\gamma \gamma \int_{\Bmin}^B \frac{B'\, dB'}{(\vect{b}\cdot\nabla B)'},
\end{equation}
which must be independent of $\chi$ due to (\ref{eq15}). A similar proof shows the separation in $\theta$ between
the points is also independent of field line.
The results $\p\Delta_\theta /\p\ang=0$ and $\p\Delta_\zeta/\p\ang=0$ are illustrated by the three arrows in Figure \ref{fig:B2}.
These arrows point along $\vect{B}$, all joining two contours of the same $B$ on opposite sides of $\Bmin$.
As $\p\Delta_\theta /\p\ang=0$ and $\p\Delta_\zeta/\p\ang=0$, these arrows must all have the same length.

It follows that the contours of $B=\Bmax$ (where $\Bmax$ is again the maximum $B$ on the flux surface)
must in fact be straight lines in the ($\theta$,$\zeta$) plane.
The basis of the argument is that as $\Delta_\zeta(\Bmax)$ and $\Delta_\theta(\Bmax)$ are constants on a flux surface,
then when the $\Bmax$ contour is translated by $\Delta_\zeta(\Bmax)$ and $\Delta_\theta(\Bmax)$,
it must lie on top of itself.  In other words, the $\Bmax$ contours
must be symmetric under a translation along $\vect{B}$,
shown by the arrow in Figure \ref{fig:B}.
Except for the uninteresting case in which $\iotab$
is a special low-order rational number, the $\Bmax$ contour cannot possibly
have this symmetry unless it is straight.
The rotational transform $\iotab$ can be assumed to be irrational, since
by continuity, $B$ on any rational surface can differ only infinitesimally from $B$ on a nearby irrational surface.
To begin the rigorous proof, first consider the $M=0$ case (poloidally closed $B$ contours), and suppose the stellarator has $\Np$ identical
toroidal segments with one $\Bmax$ contour per segment.  Let one $\Bmax$ contour be given by $\zeta = \y(\theta)$.
If we shift this contour by $\Delta_\theta(\Bmax)$ and $\Delta_\zeta(\Bmax)$, it must lie on top of the next $\Bmax$ contour,
the one given by $\zeta = \y(\theta) + (2\pi/\Np)$.  Therefore,
$\y(\theta-\iotab \Dz) + \Dz = \y(\theta) + (2\pi/\Np)$.
Then expanding $\y$ as the Fourier series $\y(\theta) = \sum_n \y_n e^{in\theta}$, we can write
\begin{equation}
\label{eq20m}
\sum_n \y_n e^{in\theta} \left( 1-e^{-i n \iotab \Dz} \right) = \Dz - \frac{2\pi}{\Np}.
\end{equation}
The $n=0$ component of this equation implies $\Dz = 2\pi/\Np$.  It follows that the exponent
$-i n \iotab \Dz$ will never be an integer multiple of $2\pi i$ if $\iotab$ is irrational.
Therefore, the quantity in parentheses in (\ref{eq20m}) can never be zero for $n \ne 0$.
Every $n\ne 0$ component of (\ref{eq20m}) consequently implies $\y_n=0$, so $\y(\theta)$
must be constant, meaning the $\Bmax$ contours are straight.
To apply the proof to fields in which both $M$ and $N$ are nonzero, all that is needed is to redefine $\y(\theta)$
as $\zeta - M\theta/N$ along the $\Bmax$ contour so $\y$ remains periodic in $\theta$.
The proof can also be adapted to the $N=0$ case by switching the roles of $\theta$ and $\zeta$.

Finally, it is important to consider both branches when forming the flux surface average
in the $(\psi,\ang,B)$ coordinate system. For any quantity $Q$, this average is given by
\begin{equation}
\label{eq21}
\left\langle Q \right\rangle =\frac{1}{V'}\sum_\gamma \gamma \int_0^{2\pi }
{d\chi }
\int_{\Bmin}^{\Bmax } {dB} \frac{Q}{\vect{B}\cdot\nabla B}
\end{equation}
where $V' = \sum_\gamma \gamma \int_0^{2\pi } {d\chi }
\int_{\Bmin}^{\Bmax} {dB} / {\rm {\bf B}}\cdot \nabla B$.

\subsection{Departure from quasisymmetry}
\label{sec:geom}

For the remainder of this section we develop several properties related to $\p h/\p\ang$,
a quantity that represents the departure from quasisymmetry.
These properties will generalize results discussed in Ref. (\cite{meQI}).
First, plugging (\ref{eq4}) into the MHD equilibrium relation $0=\nabla \psi \cdot\nabla\times\vect{B}$, we find
$\partial B_B / \partial \ang = \partial B_\chi /\partial B$.
Plugging in our earlier expressions for $B_B$ and $B_\ang$, then
\begin{equation}
\label{eq16}
(\partial / \partial \chi ) (B^2 / {\rm {\bf B}}\cdot \nabla B)
=\partial^2h / \partial B\partial \chi.
\end{equation}
Applying $\sum_\gamma \gamma $ and recalling (\ref{eq15}), then
$\sum_\gamma \gamma \, \partial^2h / \partial B\partial \chi =0$.
Now integrate this expression in $B$ from $\Bmin$ to $B$.
As $\partial h/\partial \chi$ is continuous everywhere, it is continuous at $\Bmin$, and so
it must be branch-independent at $\Bmin$. Therefore
the contribution from the integration boundary at $\Bmin$ vanishes.
Consequently,
\begin{equation}
\label{eq19}
\sum\nolimits_\gamma \gamma \, \partial h / \partial \ang=0.
\end{equation}
In other words, $\partial h/\partial \chi $ is branch-independent everywhere.

Any quantity that is continuous and branch-independent must be a constant along the curve $B=\Bmax$.
This result applies in particular to $\p h/\p\ang$, for which the constant must be zero,
or else $\int_0^{2\pi}d\ang \p h/\p\ang$ would be nonzero. Thus, $\p h/\p\ang$
must vanish along $B=\Bmax$.

\begin{figure}
\includegraphics{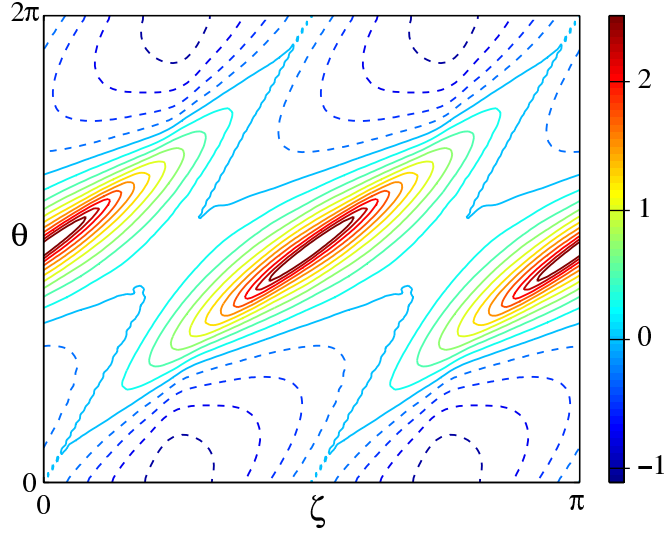}
\caption{(Color online) The departure from quasisymmetry, $\p h/\p\chi$, normalized by $G+\iotab I$,
and calculated for the field of figure \ref{fig:B} using (\ref{eq33}).
Dashed contours are negative.
\label{fig:dhdchi}}
\end{figure}

Now we derive expressions to relate the new quantity $\p h/\p\ang$ to more
familiar Boozer coordinates.
Using (\ref{eq20}),
\begin{equation}
\label{eq26}
\frac{B^2}{{\rm {\bf B}}\cdot \nabla B}=\left( {G+\iotab I} \right)\left(
{\frac{\partial B}{\partial \zeta }+\iotab \frac{\partial B}{\partial \theta
}} \right)^{-1}.
\end{equation}
In addition, using $(d\zeta /d\theta)_\ang=\iotab^{-1}$ gives
\begin{equation}
\label{eq27}
\left( {\frac{\partial B}{\partial \theta }} \right)_\chi
= {\frac{\partial B}{\partial \theta }}
 +\frac{1}{\iotab } {\frac{\partial B}{\partial \zeta }},
\end{equation}
where subscripts on partial derivatives indicate quantities held fixed. Combining this result with (\ref{eq26}),
\begin{equation}
\label{eq28}
\frac{B^2}{{\rm {\bf B}}\cdot \nabla B}
=\frac{G+\iotab
I}{\iotab } {\frac{\partial \theta }{\partial B}}
=\left(
{G+\iotab I} \right) \frac{\partial \zeta }{\partial B}
.
\end{equation}
Next, we form ${\rm {\bf B}}\times \nabla \psi \cdot \nabla B=\left( {\nabla \psi \times
\nabla \theta \cdot \nabla \zeta } \right)\left[ G(\partial B /\partial \theta )-I(\partial B/\partial \zeta ) \right]$.
Combining this result with (\ref{eq26}) gives
\begin{equation}
\label{eq30}
\frac{{\rm {\bf B}}\times \nabla \psi \cdot \nabla B}{{\rm {\bf B}}\cdot
\nabla B}
=\left( {G\frac{\partial B}{\partial \theta }-I\frac{\partial
B}{\partial \zeta }} \right)
\left( {\frac{\partial B}{\partial \zeta
}+\iotab \frac{\partial B}{\partial \theta }} \right)^{-1}.
\end{equation}
Substituting for the left-hand side using (\ref{eq13}), after some manipulation we obtain
\begin{equation}
\label{eq33}
\frac{\partial h}{\partial \chi }=-\left( {G+\iotab I} \right)\left[
{M+\left( {N-\iotab M} \right)\left( {\frac{\partial B}{\partial \zeta
}+\iotab \frac{\partial B}{\partial \theta }} \right)^{-1}\left(
{\frac{\partial B}{\partial \theta }} \right)} \right].
\end{equation}
This expression allows explicit calculation of $\p h/\p\ang$ for a given $B(\theta,\zeta)$.
As an example, Figure \ref{fig:dhdchi} shows $\p h/\p\ang$ calculated using this formula for the example
field of Figure \ref{fig:B}.

Lastly, we derive some additional formulae for $\p h/\p\ang$ that will
be used later.
From $\p \ang/\p\theta=1/(N-\iotab M)$, we find
$\p B/\p\theta = (\p B/\p\theta)_\ang + (N-\iotab M)^{-1} (\p B/\p\ang)_\theta$.
Plugging this result into (\ref{eq33}) gives
\begin{equation}
\label{eq39}
\frac{\partial h}{\partial \chi }
=-\frac{\left( {G+\iotab I} \right)}{\iotab
}\left[ {N+\frac{\left( {\partial B/\partial \chi } \right)_\theta }{\left(
{\partial B/\partial \theta } \right)_\chi }} \right]
=\frac{\left( {G+\iotab I} \right)}{\iotab
}\left( \frac{\partial \theta }{\partial \chi } -N \right),
\end{equation}
where the second equality follows from
$0 = (\p B/\p\theta)_B = (\p B/\p\theta)_\ang + (\p\ang/\p\theta)_B (\p B/\p\ang)_\theta$.
Applying $\p/\p \ang$ to (\ref{eq:ang}), we find
$\p\theta / \p\ang -N = \iotab (\p\zeta / \p\ang) - \iotab M$.
When inserted into (\ref{eq39}), this gives
\begin{equation}
\label{eq41}
\frac{\partial h}{\partial \chi }=\left( {G+\iotab I} \right)\left( {
{\frac{\partial \zeta }{\partial \chi }}  -M} \right).
\end{equation}
Comparing either (\ref{eq39}) or (\ref{eq41}) with (\ref{eq28}), it is evident that (\ref{eq16}) is
satisfied.

Expressions (\ref{eq33})(\ref{eq41}) will be used in section \ref{sec:flows}
to calculate the \PS flow and current.

\section{Parallel Flow and Current}
\label{sec:parallelTransport}

We now move on to analyzing the physical properties of omnigenous plasmas.
First we will derive general properties of the flow and current for any collisionality.
Then we will calculate the flow and current explicitly for the long-mean-free-path regime.

\subsection{Form of the flows and current}
\label{sec:flows}
Consider the density $n$ and flow velocity $\vect{V}$ of a single species.
We take $n$ to be a flux function to leading order, and take the
perpendicular flow to be given by the sum of ${\rm {\bf E}}\times {\rm {\bf
B}}$ and diamagnetic flows.
Then the flow must satisfy
$0=\nabla \cdot {\rm {\bf V}}
={\rm {\bf B}}\cdot \nabla ( V_{\vert \vert }/B)
+\omega {\rm {\bf B}}\times \nabla \psi \cdot \nabla (1/B^2)$,
where
$\omega =c (d\Phi_0/d\psi) +c (dp/d\psi) (Zen)^{-1}$, $\Phi_0 = \langle\Phi\rangle$ is the flux surface average of the
electrostatic potential $\Phi$, $p$ is the species pressure and a flux function to lowest order,
and we have used the fact that $\nabla \cdot ( {{\rm {\bf B}}\times \nabla \psi } )=0$
for MHD equilibrium.
Now define $U$ to be the single-valued and continuous solution of
\begin{equation}
\label{eq45}
{\rm {\bf B}}\cdot \nabla ( {U/B} )={\rm {\bf B}}\times \nabla\psi \cdot \nabla ( {1/B^2} )
\end{equation}
with the integration constant chosen so that $\left\langle {UB}
\right\rangle =0$. The solubility condition of (\ref{eq45}) is satisfied for any MHD
equilibrium. Then
$0={\rm {\bf B}}\cdot \nabla [ (V_{\vert \vert } +\omega U)/B ]$,
so
$V_{\vert \vert } +\omega U =A\left( \psi \right) B$
for some unknown $A$, which can be eliminated by multiplying the last equation by $B$ and flux surface averaging. Thus
\begin{equation}
\label{eq50}
V_{\vert \vert } =\frac{\langle {V_{\vert \vert } B} \rangle
B}{\left\langle {B^2} \right\rangle }-\omega U.
\end{equation}
To solve for $U$, we use (\ref{eq13}), (\ref{eq45}), and $\vect{B}\cdot\nabla = (\vect{B}\cdot\nabla B) \p / \p B$
to obtain
\begin{equation}
\label{eq52}
\frac{\partial }{\partial B}\left( \frac{U}{B}
\right)=\frac{2}{B^3}\frac{1}{\left( {N-\iotab M} \right)}\left(
{MG+NI+\frac{\partial h}{\partial \chi }} \right).
\end{equation}
Integrating,
\begin{equation}
\label{eq53}
\frac{U}{B}=Y-\frac{1}{B^2}\frac{\left( MG+NI \right)}{\left( {N-\iotab M}
\right)}-\frac{2}{\left( {N-\iotab M}
\right)}\int_B^{\Bmax} {\frac{d{B}'}{\left( {{B}'} \right)^3}}  {\frac{\partial
{h}'}{\partial \chi }}
\end{equation}
where $Y(\psi)$ is an integration constant.
The limit of integration has been chosen to ensure $Y$ is continuous, which can be seen as follows.
The parallel flow $V_{\vert \vert } $ has a term proportional to $U$ (in (\ref{eq50})), so $U$ must
be continuous everywhere. At $B=\Bmin$, then $U$ must be independent
of $\gamma $, and so from (\ref{eq53}), $Y$ then must be independent of $\gamma $
(since the other terms are all $\gamma $-independent.) Next, consider that
aside from perhaps $Y$, all the other terms in (\ref{eq53}) are continuous across
the curve
$B=\Bmax $
at fixed $\theta $ (or, in the $N=0$ case, at fixed $\zeta $.) Therefore $Y$
must have this same property. Therefore $Y$ must be independent of $\chi $.

The unknown $Y$ can be determined by multiplying (\ref{eq53}) by $B^2$ and flux surface averaging.
Defining
\begin{equation}
\label{eq56}
\W \left( {\psi ,\chi ,B}
\right)=2B^2\int_B^{\Bmax} {\frac{d{B}'}{\left( {{B}'} \right)^3}\frac{\partial {h}'}{\partial
\chi }}
\end{equation}
then
\begin{equation}
\label{eq57}
U=-\frac{1}{B\left( {N-\iotab M} \right)}\left[ {\left( MG+NI
\right)\left( {1-\frac{B^2}{\left\langle {B^2} \right\rangle }}
\right)+\W } \right].
\end{equation}

Since we constructed $U$ to satisfy $\langle UB\rangle=0$,
it should be the case that $\langle W B \rangle=0$.  Indeed, it can be verified that $\langle WB^a\rangle=0$ for any $a$ as follows.  First, using
(\ref{eq15}) and (\ref{eq19}), we find
\begin{equation}
\label{eq58}
\sum_\gamma \gamma \int_0^{2\pi}d\ang \frac{1}{\vect{B}\cdot\nabla B}\frac{\p h}{\p\ang}
= \left( \sum_\gamma \frac{\gamma}{\vect{B}\cdot\nabla B}\right) \left( \int_0^{2\pi} d\ang \frac{\p h}{\p\ang}\right) = 0
\end{equation}
where the last equality follows because the last term in parentheses is zero.
Examining the form of the flux surface average in
(\ref{eq21}), it can be seen that $\langle W B^a \rangle$ is proportional to the leftmost expression in (\ref{eq58}), so it vanishes.

Using (\ref{eq57}), the parallel flow is
\begin{equation}
\label{eq59}
V_{\vert \vert } =\frac{\langle {V_{\vert \vert } B} \rangle
B}{\langle {B^2} \rangle }+V_{\vert \vert }^{PS}
\end{equation}
where
\begin{equation}
\label{eq60}
V_{\vert \vert }^{PS} =\frac{c}{B\left( {N-\iotab M} \right)}\left(
{\frac{d\Phi_0 }{d\psi }+\frac{1}{Zen}\frac{dp}{d\psi }} \right)\left[ {\left(
{1-\frac{B^2}{\left\langle {B^2} \right\rangle }} \right)\left( MG+NI
\right)+\W } \right].
\end{equation}
Summing over species $a$ to form $j_{\vert \vert } =\sum\nolimits_a {Z_a en_a V_{\vert \vert a} } $, and using quasi-neutrality,
\begin{equation}
\label{eq61}
j_{\vert \vert } =\frac{\langle {j_{\vert \vert } B} \rangle
B}{\left\langle {B^2} \right\rangle }+j_{\vert \vert }^{PS}
\end{equation}
where
\begin{equation}
\label{eq62}
j_{\vert \vert }^{PS} =\frac{c}{B\left( {N-\iotab M} \right)}\left(
{\frac{dp_\Sigma }{d\psi }} \right)\left[ {\left( {1-\frac{B^2}{\left\langle
{B^2} \right\rangle }} \right)\left( MG+NI \right)+\W } \right]
\end{equation}
and $p_\Sigma$ is now the total pressure $\sum_a p_a$. Notice that (\ref{eq59})-(\ref{eq62}) are valid for any collisionality regime.

In a general stellarator, the \PS flow and current can only be defined implicitly, in terms of the solution
of the magnetic differential equation (\ref{eq45}). In omnigenous fields, in contrast,
the flow and current may be written explicitly, in terms of the \emph{integral} of the field strength $\W$. We now describe
two approaches to efficiently calculate $\W$ for a given $B(\theta,\zeta)$.

In the first approach, the integration variable
in (\ref{eq56}) is changed to $\theta$ using (\ref{eq27}), and then
(\ref{eq33}) is applied, giving $\W=2B^2 (G+\iotab I) w/\iotab$ ,where
\begin{equation}
\label{eq94}
w =
\int_{\thmax }^\theta {\frac{d{\theta }'}{\left( {{B}'} \right)^3}} \left(
M \frac{\partial {B}'}{\partial \zeta }  +N
{\frac{\partial {B}'}{\partial \theta }}  \right)
= \iotab \int_{\zmax }^\zeta {\frac{d{\zeta }'}{\left( {{B}'} \right)^3}} \left(
{M {\frac{\partial {B}'}{\partial \zeta }}  +N
{\frac{\partial {B}'}{\partial \theta }} } \right) .
\end{equation}
Here, $\thmax$ and $\zmax$ represent the values of $\theta$ and $\zeta$ associated with $B=\Bmax$.
Notice the integrals in (\ref{eq94}) are evaluated along constant-$\ang$ paths (field lines).

\begin{figure}
\includegraphics{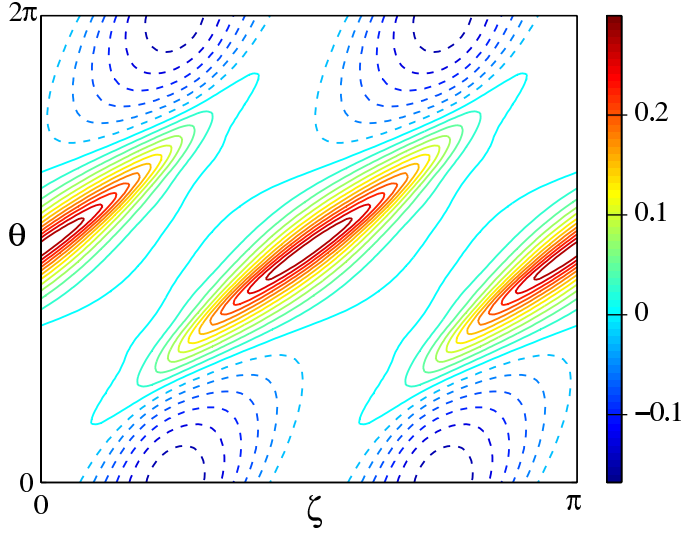}
\caption{(Color online) The quantity $W$ that
arises in the \PS flow and current, normalized by $(G+\iotab I)$, and computed for the magnetic field in Figure \ref{fig:B}.
\label{fig:W}}
\end{figure}

In an alternative approach, (\ref{eq39}) or (\ref{eq41}) is substituted into (\ref{eq56}) to obtain
\begin{equation}
\label{eq91}
 w =\iotab \int_B^{\Bmax} {\frac{d{B}'}{\left( {{B}'} \right)^3}\left( { {\frac{\partial
\zeta }{\partial \chi }}  -M} \right)}
 =\int_B^{\Bmax} {\frac{d{B}'}{\left( {{B}'} \right)^3}\left( { {\frac{\partial
\theta }{\partial \chi }}  -N} \right)} .
\end{equation}
Numerical root-finding can be used to compute $\zeta(\ang,B)$ or $\theta(\ang,B)$ from a given $B(\theta,\zeta)$,
and the result used to compute either of the integrals in (\ref{eq91}).
Figure \ref{fig:W} shows the $W$ computed by either of these methods for the
magnetic field of Figure \ref{fig:B}.

\subsection{Long-mean-free-path regime}
Explicit formulae for $\langle V_{||}B\rangle$ and $\langle j_{||} B \rangle$ can be
derived in the long-mean-free-path regime.
We begin with the drift-kinetic equation
\begin{equation}
\label{eq66}
\vpar {\rm {\bf b}}\cdot \nabla f_1 +\vm
\cdot \nabla \psi \frac{\partial f_0 }{\partial \psi }
+\frac{Ze\vpar f_0}{T}\vect{b}\cdot\nabla\Phi_1
=C\left\{ {f_1 }
\right\},
\end{equation}
where $C$ is the linearized collision operator and
$\Phi_1 = \Phi - \Phi_0$.
In (\ref{eq66}) and hereafter,
the independent velocity-space variables
are
$\lambda = \vperp^2 / (B v^2)$ and the leading-order total energy $mv^2/2+ Ze\Phi_0$.

It is useful to next define $\Delta $ by the relation
\begin{equation}
\label{eqDelta}
\vm \cdot \nabla \psi =\vpar {\rm {\bf
b}}\cdot \nabla \Delta .
\end{equation}
The radial magnetic drifts must have such a form because, from the definition of omnigenity, $\vm \cdot \nabla \psi$
vanishes upon a transit or bounce average.
Observing $\vect{B}\times\vect{\kappa}\cdot\nabla\psi = \vect{b}\times\nabla B\cdot\nabla\psi$, then
$\vm\cdot\nabla\psi = -(\vect{B}\times\nabla B \cdot\nabla \psi)(\vpar/B)\p / \p B(\vpar/ \Omega)$.
Then
(\ref{eq13}) and (\ref{eqDelta}) imply
\begin{equation}
\label{eq67}
\frac{\partial \Delta }{\partial B}=-\frac{1}{\left( {N-\iotab M}
\right)}\left( {MG+NI+\frac{\partial h}{\partial \chi }}
\right)\frac{\partial }{\partial B}\frac{\vpar }{\Omega
}.
\end{equation}
Integrating in $B$, we find
\begin{equation}
\label{eq68}
\Delta =\left( {\frac{MG+NI}{\iotab M-N}} \right)\frac{\vpar }{\Omega }+S
\end{equation}
where
\begin{equation}
\label{eq68b}
S = \frac{1}{N-\iotab M}\int_B^{B_x } {d{B}'} \frac{\partial
{h}'}{\partial \chi }\frac{\partial }{\partial {B}'}\frac{\vpar' }{{\Omega }'},
\end{equation}
and $B_x = \Bmax$ if $\lambda < 1/\Bmax$ and $B_x = 1/\lambda$ otherwise.
The limit of integration is chosen this way for passing particles ($\lambda < 1/\Bmax$)
so that $\Delta$ is continuous in position space across the curve $B=\Bmax$, and the limit of integration
for trapped particles $(\lambda>1/\Bmax)$ is chosen so that $\Delta$ is continuous in velocity space at $\lambda=1/\Bmax$.

Thus, the kinetic equation may be written $\vpar \vect{b}\cdot\nabla g = C\{ f_1\}$,
where
\begin{equation}
\label{eq69}
g=f_1 + \Delta \frac{\p f_0}{\p \psi} + \frac{Z e \Phi_1}{T} f_0.
\end{equation}
For low collisionality, we expand $g=g^{(0)} + g^{(1)} + \ldots$ and take the leading-order kinetic equation to be
$\vpar \vect{b}\cdot\nabla g^{(0)}=0$, so $\p g^{(0)}/\p B=0$. For passing particles, continuity of $g^{(0)}$ at $\Bmax$ requires $g^{(0)}$ to be constant on a flux surface,
whereas for trapped particles, $g^{(0)}$ may still depend on $\ang$.  The next-order kinetic equation is
\begin{equation}
\vpar \vect{b}\cdot\nabla g^{(1)} = C \{ g^{(0)} - \Delta \p f_0 / \p \psi \}.
\end{equation}
By annihilating $g^{(1)}$ in this equation, a constraint is obtained that determines $g^{(0)}$.
Just as in the tokamak calculation, $g^{(0)}=0$ is a solution in the trapped part of phase space.
For the passing
part of phase space, the annihilation operation is $\langle (B/\vpar) (\,\cdot\,) \rangle$.  Recalling (\ref{eq21}),
the constraint equation becomes
\begin{equation}
\label{eq71}
0= \frac{1}{V'}\sum_\gamma \gamma \int_0^{2\pi}d\ang\int_{\Bmin}^{\Bmax} dB \frac{B}{\vpar \vect{B}\cdot\nabla B} C\left\{ g^{(0)} - \Delta \frac{\p f_0}{\p \psi} \right\}.
\end{equation}

We next argue that the $S$ term in $\Delta$ (in (\ref{eq68})) may be dropped in (\ref{eq71}).
The collision operator may be written as derivatives and integrals involving $v$ and $\xi = \sigma \sqrt{1-\lambda B}$, where $\sigma = \sgn(\vpar)$,
so $C$ does not introduce any dependence on $\ang$ or $\gamma$. Thus, the contribution to $(\ref{eq71})$ from the $\p h'/\p\ang$ term in $\Delta$
is proportional to the leftmost expression in (\ref{eq58}), so it vanishes.  The constraint equation therefore reduces to
\begin{equation}
\label{eq72}
0= \left\langle\frac{B}{\vpar} C \left\{ g^{(0)} - \left( \frac{MG+NI}{\iotab M-N}\right) \frac{\vpar}{\Omega} \frac{\p f_0}{\p \psi} \right\} \right\rangle
\end{equation}
This equation is identical to the one solved for a tokamak, aside from the constant factor in parentheses.

We now show that the parallel flow associated with the distribution function $f_1$ is consistent with the form (\ref{eq59})-(\ref{eq60}) found in the previous
section from a fluid approach.  Forming $\int d^3 v\, \vpar f=n\Vpar$ using (\ref{eq69}) and taking $g \approx g^{(0)}$,
\begin{eqnarray}
\label{eq73}
n\Vpar &=& X + \frac{cn}{B} \left( \frac{d\Phi_0}{d\psi}+\frac{1}{Zen}\frac{d p}{d\psi}\right) \left[ \left( \frac{MG+NI}{N-\iotab M}\right) \right. \\
&& \left. +\frac{3B^2/4}{N-\iotab M}
\int_0^{1/B}d\lambda \int_0^{B_x}dB' \frac{\p h'}{\p\ang} \frac{(2-\lambda B')}{(B')^2 \sqrt{1-\lambda B'}} \right], \nonumber
\end{eqnarray}
where $X=\int d^3v\, \vpar g$.
Noting
\begin{equation}
\label{eq74}
\int d^3v \, Q = \pi B \sum_\sigma \sigma \int_0^{1/B} d\lambda \int_0^{\infty} dv\frac{v^3 Q}{\vpar}
\end{equation}
for any $Q$, the upper limit of the $\lambda$ integral can be changed to $1/\Bmax$
in the integral for $X$ because $g=0$ in the trapped region.  Thus, $X$ varies with position only through the factor $B$, so $X$ has the form of the $\langle \Vpar B\rangle B/\langle B^2 \rangle$
term in (\ref{eq59}).
Next, the $\lambda$ integral in (\ref{eq73}) can be evaluated by moving it inside the $B'$ integral.
In this exchange, the range of the $\lambda$ integration becomes $(0,\, 1/B')$ and the range of the $B'$ integral becomes
$(B,\, \Bmax)$. When the $\lambda$ integral is evaluated, the result is identical to the $\W$ term in (\ref{eq60}).
Consequently, the parallel flow (\ref{eq73}) evaluated from kinetic theory has precisely the spatial dependence
calculated from fluid theory in (\ref{eq59}) and (\ref{eq60}).

For the ions in a pure plasma, $g_i $ and $X$ can be calculated explicitly
using the momentum-conserving model collision operator $C_i =\nu \mathcal{L}\left\{ {f_{i1}
-f_{i0} m_i uv_{\vert \vert } /T_i } \right\}$, where
\begin{equation}
\label{eq991}
\mathcal{L}=\frac{2v_{\vert \vert } }{v^2B}\frac{\partial }{\partial \lambda }\lambda
v_{\vert \vert } \frac{\partial }{\partial \lambda }
\end{equation}
is the Lorentz pitch-angle scattering operator,
\begin{equation}
\label{eq992}
u=\left( {\int {d^3v\,f_{i0} \frac{m_i v^2}{3T_i }\nu } }
\right)^{-1}\int {d^3v\, f_{i1} \nu v_{\vert \vert } } ,
\end{equation}
\begin{equation}
\label{eq993}
\nu =\frac{2\pi Z^4e^4n_i \ln \Lambda }{\sqrt {2m_i } T_i^{3/2}
}\frac{\left[ {\mbox{erf}\left( x \right)-\Psi \left( x \right)}
\right]}{x^3},
\end{equation}
$\Psi \left( x \right)=\left[ {\mbox{erf}\left( x \right)-x\mbox{\thinspace
}\left( {d\mbox{\thinspace erf}\left( x \right)/dx} \right)} \right]/\left(
{2x^2} \right),
\quad
\mbox{erf}\left( x \right)=\left( {2/\sqrt \pi } \right)\int_0^x {\exp
\left( {-t^2} \right)dt} $ is the error function, and $x=v/\sqrt {2T_i /m_i }
$. This model operator captures the dominant effect of collisions for large aspect ratio.
The solution of (\ref{eq72}) is performed exactly as for the tokamak calculation
\cite{PerBook}, giving
\begin{equation}
\label{eq994}
g_i =f_{i0} \frac{\left( {MG+NI} \right)}{\left( {\iotab M-N}
\right)}\frac{m_i c}{ZeT_i }\frac{dT_i }{d\psi }\left( {\frac{m_i v^2}{2T_i
}-1.33} \right)\frac{\sigma v}{2}H\int_\lambda^{1/\hat{B}}
{\frac{d{\lambda }'}{\left\langle {\sqrt {1-{\lambda }'B} } \right\rangle }}
,
\end{equation}
where $H=H(\Bmax^{-1}-\lambda  )$ is a Heavyside function which is 1 for passing
particles and 0 for trapped particles. Thus, the ion parallel flow for low
collisionality is
\begin{equation}
\label{eq995}
V_{i\vert \vert } =-1.17f_c \frac{\left( {MG+NI} \right)cB}{\left( {N-\iotab
M} \right)Ze\left\langle {B^2} \right\rangle }\frac{dT_i }{d\psi
}+\frac{c}{B}\frac{\left( {MG+NI+W} \right)}{\left( {N-\iotab M}
\right)}\left( {\frac{d\Phi_0 }{d\psi }+\frac{1}{Zen_i }\frac{dp_i }{d\psi
}} \right),
\end{equation}
where
\begin{equation}
\label{eq996}
f_c =\frac{3}{4}\left\langle {B^2} \right\rangle
\int_0^{1/\Bmax} {\frac{\lambda \mbox{\thinspace }d\lambda }{\left\langle {\sqrt
{1-\lambda B} } \right\rangle }}
\end{equation}
is the effective fraction of circulating particles.

When the average parallel ion flow $\langle V_{i||} B \rangle$ is evaluated from (\ref{eq995}),
the $\W$ term - the only term that reflects the departure from quasisymmetry - does not contribute due to (\ref{eq58}).
The resulting expression for $\langle V_{i||} B \rangle$ in an omnigenous plasma
is identical to the corresponding formula for a quasisymmetric stellarator\cite{Boozer83, meQS}.
It can be seen that the bootstrap current $\langle j_{||} B \rangle$ in an omnigenous stellarator
must also be given by the same expression as in a quasisymmetric stellarator.
To understand this result, first recall that the perturbed distribution function of each
species is given by $f_1 = g^{(0)} + Ze \Phi_1 f_0 / T + \Delta \p f_0 / \p\psi$, with $\Delta$ given by
(\ref{eq68}), and $g^{(0)}$ the solution of (\ref{eq72}). The $S$ term in $\Delta$
does not contribute to $\langle V_{||}B \rangle$ for the species as we have seen due to (\ref{eq58}).
Also, $g^{(0)}$ must be the same as in a quasisymmetric stellarator, since in the equation that determines it, (\ref{eq72}),
the departure from quasisymmetry does not appear.  Thus, $\langle V_{||}B \rangle$ for each species is
the same as in a quasisymmetric device, and so $\langle j_{||} B \rangle$ is the same as well.
The result is
\begin{equation}
\label{eqb1}
\left\langle {j_{\vert \vert } B} \right\rangle =
\frac{f_t ca}{Z\left( {\sqrt 2 +Z} \right)}
\frac{(MG+NI)}{(N-\iotab M)}\left( {\frac{dp_i
}{d\psi }+\frac{dp_e }{d\psi }-\frac{2.07Z+0.88}{a}n_e
\frac{dT_e }{d\psi }-1.17\frac{n_e }{Z}\frac{dT_i }{d\psi }} \right),
\end{equation}
where $f_t =1-f_c $ is the effective trapped fraction, $a=Z^2+2.21Z+0.75$, and $Z$ is the ion charge. To obtain (\ref{eqb1}), the method of page 207 of
Ref. \onlinecite{PerBook} can be employed, using the approximate Spitzer function from appendix B
of Ref. \onlinecite{Istvan} with two Laguerre polynomials.
The tokamak expressions for $\langle V_{||} B \rangle$ and $\langle j_{||} B \rangle$ can be recovered from the omnigenous/quasisymmetric stellarator
expressions by simply setting $N=0$.

The expression (\ref{eqb1}) can also be derived by solving the differential equations
for the bootstrap current in a general stellarator, derived using the Shaing-Callen
moment approach\cite{ShaingCallen, Wakatani, Johnson}.  In this approach, the bootstrap current in a general stellarator
for a pure $Z=1$ plasma is given by
\begin{equation}
\label{eqb2}
\langle j_{||} B \rangle = -1.70 c\frac{f_t}{f_c} \Gbs \left( \frac{dp_e}{d\psi}+\frac{dp_i}{d\psi} - 0.75 n\frac{dT_e}{d\psi} - 1.17 n \frac{dT_i}{d\psi}\right)
\end{equation}
where the ``geometric factor" is
\begin{equation}
\label{eqb3}
\Gbs = \frac{1}{f_t} \left( \langle g_2 \rangle - \frac{3}{4}\langle B^2 \rangle \int_0^{1/\Bmax} \lambda \frac{\langle g_4 \rangle}{\langle g_1 \rangle} d\lambda \right),
\end{equation}
$g_1 = \sqrt{1-\lambda B}$, $g_2$ is the solution of
\begin{equation}
\label{eqb4}
\vect{B}\cdot\nabla (g_2/B^2 ) = \vect{B}\times\nabla\psi\cdot\nabla (1/B^2)
\end{equation}
such that $g_2=0$ when $B=\Bmax$, and $g_4$ is the solution of
\begin{equation}
\label{eqb5}
\vect{B}\cdot\nabla (g_4/g_1) = \vect{B}\times\nabla\psi\cdot\nabla (1/g_1)
\end{equation}
for any $\lambda$ in the range $[0,\, 1/\Bmax]$ such that $g_4=0$ when $B=\Bmax$.
Equation (\ref{eqb4}) for $g_2$ closely resembles equation (\ref{eq45}) for $U$ which we solved earlier.
Using the solution (\ref{eq57}), then
\begin{equation}
\label{eqb6}
g_2 = \frac{1}{N-\iotab M} \left[ (MG+NI) \left( \frac{B^2}{\Bmax^2} -1 \right) -\W \right].
\end{equation}
Equation (\ref{eqb5}) for $g_4$ may be solved in the same manner used to find $U$ in (\ref{eq45})-(\ref{eq57}), yielding
\begin{equation}
\label{eqb7}
g_4 = \frac{MG+NI}{N-\iotab M} \left[ \frac{\sqrt{1-\lambda B}}{\sqrt{1-\lambda \Bmax}} -1 + \frac{\lambda}{2}\sqrt{1-\lambda B} \int_B^{\Bmax} \frac{dB'}{(1-\lambda B')^{3/2}} \frac{\p h'}{\p\ang} \right].
\end{equation}
Evaluating (\ref{eqb3}), the $\W$ term in $g_2$ and the $\p h'/\p\ang$ term in $g_4$ vanish
upon flux surface averaging due to (\ref{eq58}).
Then evaluating the $\lambda$ integral, we obtain $\Gbs = (MG+NI)/(\iotab M-N)$, which is precisely
the result for a quasisymmetric stellarator.  Then (\ref{eqb2}) for $f_t \ll 1$ reduces to the $Z=1$ limit of (\ref{eqb1}), proving the two approaches are consistent.

Equation (\ref{eqb1}) gives the current \emph{density} on one flux surface in terms of $I$ and $G$,
which represent the \emph{total} poloidal and toroidal current through a suitable surface (times $2/c$).
By writing $I$ and $G$ as the appropriate
integrals of the current density, as shown in Appendix \ref{a:current}, two
ordinary differential equations (\ref{aj:eq6m}) and (\ref{aj:eq7}) can be derived
that give $I$ and $G$ in terms of $\langle j_{||} B \rangle$.
These two equations, together with (\ref{eqb1}), constitute a coupled
system of equations for the self-consistent current profile.

The case
$M=0$ (generalized quasi-poloidal symmetry) is noteworthy,
for then the substitution of (\ref{eqb1}) into (\ref{aj:eq6m}) gives
$d I/d\psi = (\ldots) I$.  As the boundary condition for $I$ is that it vanishes
on the magnetic axis, then the self-consistent profile of $I(\psi)$ is $I=0$,
and so $\langle j_{||}B\rangle=0$ everywhere.  Thus, we recover the result
of Refs. \onlinecite{Subbotin}-\onlinecite{meQI} that the bootstrap current in an $M=0$ omnigenous
device vanishes.  (If Ohmic or RF-driven current is present, there will be additional
contributions to $\langle j_{||} B \rangle$ besides (\ref{eqb1}),
providing an inhomogeneous term in the differential equation for $I(\psi)$,
so $I$ and the bootstrap current would become nonzero.)
In contrast, for any omnigenous plasma with $M \ne 0$, the contribution from
$G$ to the bootstrap current is still present. As $G$ has a nonzero boundary condition at the plasma edge, then $G$ will generally be nonzero, and so the bootstrap current will also be nonzero.

\section{Radial electric field}
\label{sec:Er}

In a non-quasisymmetric stellarator, there is usually only one special value of the radial electric field $\Er$ at each radius
for which the electron and ion particle fluxes will be equal.
In equilibrium, $\Er$ must therefore take on this value.
(If the electron temperature $T_e$ is much higher than the ion temperature $T_i$, a second ``electron root" solution may also be possible,
but we will assume $T_e \sim T_i$, excluding this possibility.)
In this section we will determine $\Er$
for the case of ions in the long-mean-free-path regime.

We begin by finding the radial particle flux of each species,
allowing general collisionality for the moment.
By applying $\langle B^{-2}\vect{B}\times\nabla\psi\cdot(\ldots)\rangle$ to the fluid momentum equation with
a diagonal pressure tensor,
the particle flux is found to be
$\left \langle \vect{\Gamma}\cdot\nabla\psi \right\rangle
= \Gamma_{m} + \Gamma_E + \Gamma_{cl}$,
where $\Gamma_{m} = \left\langle \int d^3v\, f \vm\cdot\nabla\psi \right\rangle$,
 $\Gamma_E = \langle n c B^{-2} \vect{B}\times\nabla\Phi\cdot\nabla\psi\rangle$
and $\Gamma_{cl} = c\langle B^{-2} \vect{B}\times\nabla\psi\cdot\int d^3v\, \vect{v}C\{f\}\rangle /(Ze)$
is the classical flux, which we henceforth neglect.
The earlier definition (\ref{eqDelta})
with (\ref{eq68}) may be substituted into $\Gamma_{m}$.
Using (\ref{eq74}), the fact that $\langle \vect{B}\cdot\nabla Q \rangle =0$ for any  single-valued $Q$,
and the fact that $\Delta=0$ at $\lambda = 1/B$, we derive
$
\Gamma_{m}
= -\left\langle \int d^3v\, \Delta \vpar  \vect{b}\cdot\nabla f \right\rangle.
$
We then substitute in the drift-kinetic equation (\ref{eq66}).
The resulting $\p f_0 / \p\psi$ term vanishes due to $\sigma$ parity, and the
$\Phi_1$ term can be shown to cancel the earlier $\Gamma_E$ term in the total flux; this can be done by again applying (\ref{eqDelta}) and
noting $\langle \vect{B}\times\nabla\psi\cdot\nabla(\Phi_1/B^2)\rangle =
\langle \nabla\cdot (\Phi_1 B^{-2} \vect{B}\times\nabla\psi)\rangle = 0$. We thereby obtain
\begin{equation}
\left \langle \vect{\Gamma}\cdot\nabla\psi \right\rangle
\approx
-\left \langle \int d^3v\, \Delta C\{f\}\right\rangle.
\end{equation}
This result holds for any collisionality regime.

Next, we flux-surface-average the quasineutrality equation $\nabla\cdot\vect{j}=0$ to obtain
$0=\langle \vect{j}\cdot\nabla\psi\rangle = \sum_a Z_a e \langle \vect{\Gamma}_a\cdot\nabla\psi\rangle$,
where $a$ is the particle species.
We consider a pure plasma with ion charge $Z$ and comparable electron and ion temperatures.  In this case, the ion
contribution to the radial current is larger than the electron contribution
by $\sim \sqrt{m_i/m_e}$,
so the leading-order ambipolarity constraint is $\langle\vect{\Gamma}_i \cdot\nabla\psi\rangle=0$.
We henceforth assume all quantities refer to ions and drop the subscripts.

Specializing to the long-mean-free-path regime, we may apply the model collision operator
from (\ref{eq991})-(\ref{eq993}), appropriate for large aspect ratio.  Using the momentum conservation property $\int d^3v\,\vpar C\{f\}=0$,
\begin{equation}
\label{eq100}
\langle\vect{\Gamma}\cdot\nabla\psi\rangle
= -\left\langle \int d^3v\, S \nu \mathcal{L} \left\{ g - \Delta \frac{\p f_0}{\p \psi} - f_0 \frac{m u \vpar}{T} \right\} \right\rangle.
\end{equation}
We now exploit the fact that $\langle Q \, \p h/\p\ang\rangle=0$ for any $Q$ that is independent of
branch and $\ang$.  This fact follows from (\ref{eq21}) and (\ref{eq58}).  Thus, only the terms in
(\ref{eq100}) that are quadratic rather than linear in $\p h/\p\ang$ will contribute.
For instance, $g$ will not contribute since it is constant on a flux surface.  Keeping only the nonvanishing terms,
we may write $\langle \Gamma\cdot\nabla\psi \rangle  = \Gamma_1 + \Gamma_2$ where
\begin{equation}
\label{eq101}
\Gamma_1 = \left\langle \int d^3v\, S \nu \mathcal{L}\left\{S \frac{\p f_0}{\p\psi}\right\}\right\rangle,
\end{equation}
\begin{equation}
\label{eq102}
\Gamma_2 = \left\langle \int d^3v\, S \nu \mathcal{L}\left\{ f_0 \frac{m \vpar u_{\ang}}{T} \right\}\right\rangle,
\end{equation}
and
\begin{equation}
\label{eq103}
u_{\ang} = -\left( \int d^3v\, f_0 \nu \frac{m v^2}{3T} \right)^{-1}
\int d^3v\, \frac{\p f}{\p\psi} \nu \vpar S
\end{equation}
is the part of $u$ that depends on $\ang$.
Applying (\ref{eq74}), it can be seen that
$\Gamma_1$ and $u_{\ang}$ are both proportional to the integral
$\propto \int_0^{\infty} dv\, v^4 \nu \p f_0/\p\psi$.
Therefore, $\Gamma_2$ and the total radial flux are proportional to the same factor.
The ambipolarity condition can thus be written
\begin{equation}
\label{eq104}
0 = \langle \Gamma\cdot\nabla\psi \rangle \propto
\left( \int_0^{2\pi} d\ang \frac{\p h'}{\p\ang} \frac{\p h''}{\p\ang} \right)
\int_0^{\infty}dv\, v^4 \nu \frac{\p f_0}{\p\psi}
\end{equation}
where the single and double primes refer to the integration variables in the $S$ factors.
In a quasisymmetric or axisymmetric plasma, $\p h/\p\ang=0$ and so the equation is
automatically satisfied.  However, in a non-quasisymmetric omnigenous field,
the $\ang$ integral in (\ref{eq104}) is generally nonzero, so
the $v$ integral following it must vanish.  This condition may be written
\begin{equation}
\label{eq105}
0=\int_0^{\infty} dx\, x^4 e^{-x^2} \nu \left[ \frac{1}{p} \frac{dp}{d\psi} + \frac{Ze}{T} \frac{d\Phi_0}{d\psi} + \left( x^2 - \frac{5}{2}\right) \frac{1}{T}\frac{dT}{d\psi}\right]
\end{equation}
which may be solved for the radial electric field $d\Phi_0/d\psi$.  Reinstating the species
subscripts for completeness, the result is
\begin{equation}
\label{eq106}
\frac{d\Phi_0}{d\psi} = \frac{T_i}{Ze} \left( -\frac{1}{p_i}\frac{dp_i}{d\psi} + \frac{1.17}{T_i} \frac{dT_i}{d\psi}\right)
\end{equation}
where
\begin{equation}
1.17 = \frac{5}{2} - \left( \int_0^{\infty}dx\, x^4 e^{-x^2}\nu \right)^{-1} \int_0^{\infty} dx\, x^6 e^{-x^2} \nu
\end{equation}
is the same numerical constant that appears in the banana-regime tokamak ion flow
and in (\ref{eq995}).  This electric field differs from previously known results for a non-omnigenous stellarator.
For example, if the main ions in a non-omnigenous device are in the $1/\nu$ regime of collisionality,
the relation (\ref{eq106}) holds but with 1.17 replaced by 2.37 (see e.g. equation (35) of Ref. \onlinecite{ConnorHastie}).

Assuming the temperature scale length is not dramatically shorter than the density scale length, (\ref{eq106})
gives an inward electric field.  Physically, this field
develops to electrostatically confine the ions, reducing their radial flux to the much smaller level of the electron flux.

As the departure from omnigenity increases, (\ref{eq106}) will become inapplicable before
the formulae for the flow and current do, since the particles with nonzero
average radial drift will contribute strongly to the radial particle
flux but not to the parallel transport.

\section{Constructing stellarator-symmetric omnigenous fields}
\label{sec:construction}
In this section we will give a construction for $B(\theta,\zeta)$ field
strength patterns that both satisfy all the omnigenity conditions
and also satisfy \emph{stellarator symmetry}\cite{sugama11}, meaning the invariance
of $B$ under the replacements $\theta\to -\theta$, $\zeta \to -\zeta$ .
A construction of omnigenous $B(\theta,\zeta)$ was given previously in
Ref. \onlinecite{CSPoP}, but that procedure generally produces fields without stellarator
symmetry, whereas every stellarator experiment to our knowledge does
possess stellarator symmetry (aside from small error fields).

The fields we will construct are designed to have the following omnigenity properties:
1) The $B$ contours will all link the torus,
2) the $\Bmax$ contour will be straight, and
3) $\p \Delta_\zeta / \p\ang=0$ and $\p\Delta_\theta /\p\ang=0$. Here,
$\Delta_\zeta$ and $\Delta_\theta$ are the same quantities discussed following (\ref{eq20b}): the separations in $\zeta$ and $\theta$ between the pair
of points on opposite branches of a field line but at the same $B$.
The other omnigenity properties from section \ref{sec:BProperties} will then follow automatically.
For example, by applying $\p^2/\p B \,\p\ang$ to (\ref{eq20b}),
then (\ref{eq15}) follows.

We will now present the construction, and we will verify afterward that it indeed produces fields
that are omnigenous and stellarator-symmetric.
The construction is given in terms of new angles $\thetah$ and $\zetah$,
such that consecutive maxima of $B$ lie on the constant-$\thetah$ curves $\zetah=0$ and $\zetah=2\pi$.
We also employ an effective rotational transform $\iotah$, equal to $d\thetah / d\zetah$ along the field.
To construct an omnigenous field with
nonzero $N$ and with $\Np N$ toroidal periods, the new quantities are related
to the original quantities by $\thetah=\theta$, $\zetah = (N\zeta-M\theta)\Np$, and $\iotah=\iotab /[(N-\iotab M)\Np]$.
To construct an omnigenous field with $(M,N)=(1,0)$ and $\Np$ toroidal periods, then instead
$\thetah=\Np\zeta$, $\zetah = \theta$, and $\iotah=\Np / \iotab$.
In either case, a stellarator-symmetric $B$ is one invariant under $\thetah\to - \thetah$ and $\zetah \to - \zetah$.
Also, $\p \Delta_\zeta /\p\ang=0$ is equivalent to $\p \Delta_{\zetah} /\p\ang=0$,
where $\Delta_{\zetah}$ is defined just as for $\Delta_{\zeta}$ but using $\zetah$ in place of $\zeta$.

There are several
inputs to the construction.  First, we may pick $\Bmin$ and $\Bmax$.
Second, we choose a function $\D(x)$ which will turn out to be closely related to $\Delta_{\zetah}(B)$.
The function $\D(x)$ must be defined on the domain $[0,\pi]$ and must satisfy $\D(0)=\pi$ and $\D(\pi)=0$.
Lastly,
we choose a function $s(x,y)$ which will represent the $\zetah$-variation of the $B$ contours.
We require $s(x,y)$ to be both odd in $y$ and $2\pi$-periodic $y$ (i.e. the Fourier series for $s$ contains
$\sin(ny)$ terms but no $\cos(ny)$ terms or $y$-independent term.)
The input $x$ ranges over $[0,\pi]$, and we require $s(0,y)=0$ for all $y$.

It is then useful to introduce a new coordinate $\eta$ which resembles
$B$ but which is different in the two branches.  Specifically, we define $\eta \in [0,2\pi]$
through the relation
\begin{equation}
\label{constr}
B=\Bmin(1+\epsilon + \epsilon \cos\eta)
\end{equation}
where $\epsilon = (\Bmax-\Bmin)/(2\Bmin)$.
Notice from (\ref{constr}) that $\eta=0$ along $\zetah=0$ (where $B=\Bmax$) and
$\eta=2\pi$ along $\zetah=2\pi$ (where $B$ again rises to $\Bmax$).
While $B$ contours coincide with $\eta$ contours in the $(\thetah, \zetah$) plane,
$\eta$ lies in the range $0\le \eta < \pi$ on one branch, while
$\eta$ lies in the range $ \pi < \eta \le 2\pi$ on the other branch.

We then compute $\zetah$ as follows:
\begin{equation}
\label{eq:zetaOfEta}
\zetah(\eta,\thetah) = \left\{
\begin{array}{ll}
\pi - s\left(\eta,\, \thetah + \iotah \D(\eta)\right) - \D(\eta)  & \mbox{if } 0\le \eta\le\pi \\
\pi + s\left(2\pi-\eta,\, -\thetah + \iotah \D(2\pi-\eta)\right) + \D(2\pi-\eta) & \mbox{if } \pi < \eta \le 2\pi.
\end{array}
\right.
\end{equation}
Numerical root-finding is next used to compute the inverse map $\eta(\thetah, \zetah)$,
and finally $B$ is calculated by (\ref{constr}).

\begin{figure}
\includegraphics{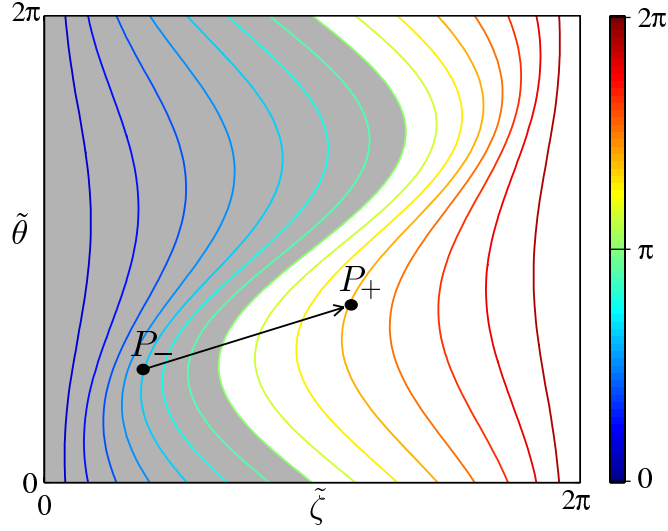}
\caption{(Color online) Contours of $\eta(\thetah, \, \zetah)$ for the field of Figure \ref{fig:B}.
One branch ($\eta<\pi$) is shaded
while the other branch ($\eta > \pi$) is unshaded.
Field lines are parallel to the arrow.  $P_-$ and $P_+$ are defined preceding (\ref{eq201}).
\label{fig:eta}}
\end{figure}

Figures \ref{fig:B} and \ref{fig:B2} show omnigenous stellarator-symmetric fields constructed using the above procedure.
For Figure \ref{fig:B}, the parameters are chosen to resemble HSX\cite{HSX}: $M=1$, $N=4$, $\Np=1$, $\iotab=1.05$, and $\epsilon=0.072$.
The numerical functions used are $s(x,y) = 0.4 x \sin(y)$ and $\D(x) = \pi-x$.
Figure \ref{fig:eta} shows the associated $\eta(\thetah,\, \zetah)$ function.
For Figure \ref{fig:B2}, $M=1$, $N=0$, $\Np=3$, $\iotab=1.62$, $\epsilon=0.1$, $s(x,y) = 0.15 x \sin(y)$ and $\D(x) = \pi-x$.

We now prove that the $B$ resulting from the above construction is stellarator-symmetric and omnigenous,
beginning with stellarator symmetry.
Thinking of $B$ as an independent variable in place of $\zetah$, we can restate the definition of this symmetry as follows: when $\thetah \to -\thetah$,
and $B$ remains constant, and the branch is reversed, then $\zetah$ must go to $-\zetah$.
Reversing the branch at constant $B$ is equivalent to the replacement $\eta \to 2\pi-\eta$.
Adding factors of $2\pi$ to keep $\thetah$ and $\zetah$ within the range $[0,\, 2\pi]$,
we can therefore write the stellarator symmetry criterion as
\begin{equation}
\label{inverseSS}
\zetah(\eta,\thetah) = 2\pi - \zetah(2\pi-\eta,\, 2\pi-\thetah).
\end{equation}
If $\eta \le \pi$, then the left-hand side of this equation is given by the top line of (\ref{eq:zetaOfEta}),
and the right-hand side of (\ref{inverseSS}) is given by the bottom line of (\ref{eq:zetaOfEta}).
It can be immediately verified that (\ref{inverseSS}) is indeed satisfied.  Similarly,
if $\eta > \pi$, then the left-hand side of (\ref{inverseSS}) is given by the bottom line of (\ref{eq:zetaOfEta}),
while the right-hand side of (\ref{inverseSS}) is given by the top line of (\ref{eq:zetaOfEta}),
and the satisfaction of (\ref{inverseSS}) is immediate.

It remains to show that $\Delta _{\zetah }$ is independent of field
line. To see this, consider two points $P_+$ and $P_-$ on the same field line and at the
same $B$ but on opposite sides of $\Bmin$. Figure \ref{fig:eta} shows two such points.
Let
$\left( \thetah_+ ,\, \zetah_+ ,\,\eta_+ \right)$ describe the point $P_+$ with $\eta
>\pi $ and let $\left( \thetah_- ,\, \zetah_- ,\, \eta_-  \right)$ describe the point $P_-$ with $\eta
<\pi $.
 Applying (\ref{eq:zetaOfEta}) to $P_-$,
\begin{equation}
\label{eq201}
 \zetah_- =\pi -s\left( {\eta_-
,\, \thetah_- +\iotah \D \left( {\eta_- }
\right)} \right)-\D \left( {\eta_- } \right).
\end{equation}
Applying (\ref{eq:zetaOfEta}) to $P_+$, noting $\eta_+ =2\pi -\eta_- $ and $\thetah_+ = \thetah_- + \iotah (\zetah_+ - \zetah_-)$,
and recalling $s$ is odd in its second input, then
\begin{equation}
\label{eq202}
\zetah_+ =\pi - s\left( \eta_- ,\, \thetah_- +\iotah (\zetah_+ - \zetah_-) -\iotah \D ( \eta_-)\right)+\D (\eta_-).
\end{equation}
Comparing (\ref{eq201}) and (\ref{eq202}),
it can be seen that a consistent solution of the system is $\zetah_+ - \zetah_- = 2 \D(\eta_-)$.
Therefore $\Delta_{\zetah} = \zetah_+ - \zetah_-$
is independent of $\thetah
$ for each $B$, so the magnetic field is omnigenous.

Given that $\iotab$ is rarely much larger than 1 in experiments, it is difficult to construct
$N=0$ omnigenous fields that depart strongly from quasisymmetry.  This is because the $B$ contours in a $N=0$
quasisymmetric field are already nearly parallel to the field lines when $\iotab / \Np <1$.  Even a slight curvature
of the $B$ contours would result in points of tangency to the field lines, and as discussed in Section \ref{sec:helicity},
the $B$ contours and field lines can never be tangent in omnigenous fields.

\section{Conclusions}

The limit of a single-helicity (i.e. quasisymmetric) field is a useful
point of reference for insight into stellarator physics, for in these effectively 2D fields,
analytic calculations can be carried out more completely than in a general stellarator.  In the preceding
sections, we have shown that omnigenous plasmas ought to be another such point of reference.
Many geometrical and physical properties of omnigenous plasmas are either identical to the associated quantity in a quasisymmetric
plasma, or else obtained by the addition of one term that is strongly constrained.
While every quasisymmetric field is omnigenous, in practice not every omnigenous field is quasisymmetric.
Therefore a broader attention to the larger class of omnigenous fields
may allow better optimization for other criteria such as smaller aspect ratio or coil simplicity.
Furthermore, it is omnigenity and not quasisymmetry that is the necessary condition for confinement
of alpha particles.
Any viable fusion reactor will need to be nearly omnigenous in order to prevent damage
to the first wall from unconfined alphas.  Indeed, alpha particle confinement time is routinely
used as an optimization criterion in stellarator design codes.  As alpha confinement
becomes increasingly important in future reactor-relevant experiments,
stellarator designs can be expected to more closely approach omnigenity.
The recent design study in Ref. \onlinecite{Subbotin}
gives one example of a nearly omnigenous device.

In Section \ref{sec:BProperties}, novel proofs were given of various geometric properties of omnigenous magnetic fields.
The $B$ contours must link the flux surface toroidally, poloidally, or both, so no isolated extrema of $B$ on the
flux surface are allowed.  The field lines can never be tangent to the $B$ contours.  The curve of $\Bmax$ (the maximum of $B$) must be straight in Boozer coordinates.
The variation of $B$ on the two branches is constrained by (\ref{eq15}).
The integers $M$ and $N$ usually defined by the quasisymmetry relation $B=B(M\theta-N\zeta)$ may be generalized
to any omnigenous field by redefining them as the number of times each $B$ contour encircles the plasma toroidally and poloidally.
In (\ref{eq13}), it was shown that the ratio $(\vect{B}\times\nabla B\cdot\nabla\psi)/\vect{B}\cdot\nabla B$,
which arises repeatedly in transport calculations,
may be written as a sum of a quasisymmetric part
and a departure from quasisymmetry, $\p h/\p\ang$.

This same division into quasisymmetric and non-quasisymmetric components also arises in physical quantities.
The \PS flow and current (\ref{eq59}) and (\ref{eq62}) are given by the same formulae as in a quasisymmetric device, with the addition of the $\W$ term,
which is an integral of $B$ along field lines.  In contrast, the \PS flow and current in a general stellarator can only be expressed
in terms of the solution of a partial differential equation like (\ref{eq45}).
The distribution function in a low-collisionality omnigenous plasma is given by the distribution function in a quasisymmetric plasma of the same helicity,
plus the term $-S \,\p f_0/\p\psi$.
This term does not contribute to the average parallel flow $\langle V_{i||}B \rangle$ and bootstrap current $\langle j_{||} B \rangle$,
so these quantities are given by exactly the same expressions as in a quasisymmetric plasma.
For each of the quantities discussed above, formulae for a quasisymmetric plasma can be recovered by setting $\p h/\p\ang \to 0$,
and tokamak formulae can be recovered by also setting $N\to 0$ and $M\to 1$.

In general, a self-consistent current profile may be found by calculating the current density in term of the total current using kinetic
theory, and then writing the current density as the appropriate derivative (\ref{aj:eq6m})-(\ref{aj:eq7}) of the total current.
When $M=0$, corresponding to generalized quasi-poloidal symmetry, the resulting self-consistent configuration has no bootstrap current.
Such configurations have several desirable properties owing to the minimization of current driven by pressure.
The magnetic field would remain omnigenous over a larger range of plasma
pressure.  Also, the equilibrium $\beta$ limit may be less severe and current-driven instability may be reduced.

The property of omnigenous plasmas that is most fundamentally different from quasisymmetric plasmas is the radial electric field.
Omnigenous plasmas that are not quasisymmetric are not intrinsically ambipolar.
The radial electric field $\Er$
is therefore determined by low-order neoclassical physics, so it is possible to solve for $\Er$
using the ambipolarity condition.  The closed-form expression (\ref{eq106}) gives the resulting radial electric field for low collisionality.
The formula is independent of the details of the magnetic field geometry.

In every stellarator experiment to date, detailed numerical neoclassical calculations\cite{BeidlerBigBenchmarking} find
a $1/\nu$ regime of radial transport to exist,
indicating that departures from omnigenity are non-negligible for radial transport in realistic experiments.  Thus,
the results in Section \ref{sec:Er} are likely to be applicable only for benchmarking numerical codes,
since in codes the $B(\theta,\zeta)$ pattern can be made more perfectly omnigenous than in a true 3D equilibrium.
However, unlike the radial transport, the parallel flow and current are carried by the bulk of the distribution function,
not by a small fraction of trapped particles.  Therefore, the formulae derived in this paper for the flow and current are robust
to small departures from omnigenity.  This resilience is aided by the fact that any $1/\nu$ part of the distribution function would be even in $\sgn(\vpar)$
and so it would carry no flow.

In conclusion, omnigenity is a useful ideal limit for gaining insight and benchmarking codes:
omnigenity is experimentally relevant,
it is more inclusive than quasisymmetry,
and yet it allows explicit analytic results to be derived that are nearly as concise as
results for quasisymmetric or axisymmetric plasmas.


%
%

%

\begin{acknowledgments}
This work was supported by the US Department of Energy
through grant DE-FG02-91ER-54109 and through the Fusion Energy Postdoctoral Research Program
administered by the Oak Ridge Institute for Science and Education.
\end{acknowledgments}

\appendix

\section{Additional Proofs}
\label{a:proofs}

In this appendix we will first prove the equivalence of the two definitions of
omnigenity: 1) the bounce-averaged radial drift vanishes, and 2) the longitudinal
adiabatic invariant is a flux function.  Then, we will derive
(\ref{eq15}).

Begin by considering a general (not necessarily omnigenous) stellarator equilibrium.
We then write
$\vm\cdot\nabla\psi
=(\vpar/\Omega)\nabla\times(\vpar \vect{b})\cdot\nabla\psi
=(\vv_{||}/\Omega)\nabla\cdot\left[ (\vv_{||}/B)\vect{B}\times\nabla\psi\right]$,
where the gradients are evaluated at fixed $\lambda$ and $v$.
The formula for the divergence in a general coordinate system is then applied, using
the coordinates $(\psi,\aang,\zeta)$, where
$\aang = \theta-\iotab\zeta$ is a field line label, and $\theta$ and $\zeta$ are any straight-field-line
poloidal and toroidal angles (so $\vect{B}=\nabla\psi\times\nabla\aang$).
Noting that the inverse Jacobian is $1/\sqrt{g}=\vect{B}\cdot\nabla\zeta$ and that $\vect{B}\times\nabla\psi\cdot\nabla\zeta = -I/\sqrt{g}$,
we thereby obtain
\begin{equation}
\label{eq:QIa0}
\vm\cdot\nabla\psi =
\frac{mc}{Ze}\vv_{||} \vect{b}\cdot\nabla\zeta
\left[ \left(\frac{\partial}{\partial \aang}\right)_{\zeta} \left( \frac{\vv_{||}}{\vect{b}\cdot\nabla\zeta} \right)
-\left(\frac{\partial}{\partial \zeta}\right)_{\aang} \left( \frac{I \vv_{||}}{B} \right) \right].
\end{equation}
Next, we define the bounce average, which for any quantity $A$ is
$\bar A =  \tau^{-1}\sum_{\sigma} \sigma \int_{\zeta_-}^{\zeta_+} (\vpar \vect{b}\cdot\nabla\zeta)^{-1} A\, d\zeta$,
where
$\tau = 2 \int_{\zeta_-}^{\zeta_+} (|\vpar | \vect{b}\cdot\nabla\zeta ) ^{-1} d\zeta$,
$\sigma=\mathrm{sgn}(\vv_{||})$, and $\zeta_-$ and $\zeta_+$ are the two bounce points (at which $\vv_{||}=0$).
The integrals in the bounce average are performed at constant $\aang$.
Applying the bounce average to (\ref{eq:QIa0}) gives
$\overline{\vm\cdot\nabla\psi} =
mc(Ze\tau)^{-1}
\sum_{\sigma} \sigma \int_{\zeta_-}^{\zeta_+} d\zeta \left(\partial / \partial\aang\right)_{\zeta} \left[ \vpar (\vect{b}\cdot\nabla\zeta)^{-1}\right]$.
Even though $\zeta_-$ and $\zeta_+$ depend on $\aang$, it is valid to pull the $\partial/\partial\aang$ derivative
in front of the integral in this result because the integrand vanishes at these endpoints.
Now, consider the longitudinal invariant for trapped particles:
$J\left( \psi,\aang ,\vv,\lambda  \right)=\oint {\vv_{\vert \vert }\,d\ell }$, where the integration is again carried out along a full bounce. Using $d\ell = d\zeta/\vect{b}\cdot\nabla\zeta$ in this definition,
then we obtain
\begin{equation}
\overline{\vm\cdot\nabla\psi} =
\frac{mc}{Ze\tau} \frac{\partial J}{\partial \alpha}.
\end{equation}
Consequently, the bounce-averaged radial drift vanishes if and only if the longitudinal invariant $J$
is a flux function, and so the two definitions of omnigenity are equivalent.

Now we move on to the proof that $(\p / \p \ang) \sum_\gamma \gamma/\vect{b}\cdot\nabla B=0$ in an omnigenous field \cite{CSPoP}.
We first note that
$J=2 \vv
\sum_\gamma \gamma \int_{\Bmin}^{1/\lambda } (\vect{b}\cdot \nabla B)^{-1}\sqrt {1-\lambda B} \,dB $.
Applying $\partial /\partial \ang $ and requiring $J$ to be constant on a
flux surface gives
\begin{equation}
\label{eq96}
0=\int_{\Bmin}^x {dB\sqrt {x-B} \mbox{\thinspace }\gCS } \left( B
\right)
\end{equation}
where $x=1/\lambda $ and
$\gCS\left( B \right)=(\partial / \partial \ang) \sum_\gamma \gamma / (\vect{b}\cdot \nabla B)$.
Equation (\ref{eq96}) is true for any $x$ in the interval $(\Bmin,\,\Bmax)$.
We now divide (\ref{eq96}) by $\sqrt {y-x} $, where $y$ is any
value in the same interval, and we integrate over $x$ from $\Bmin$ to $y$. Interchanging
the order of the $x$ and $B$ integration, the $x$ integral becomes
$\int_B^y {dx} \sqrt{ (x-B)/(y-x)} =(\pi/2)\left( {y-B} \right)$,
giving
$0=\int_{\Bmin}^y dB\,\left( {y-B} \right)\gCS
\left( B \right)$.
Differentiating twice with respect to $y$ gives
$0=\gCS\left( y \right)$.
We had allowed $y$ to be any value in the interval $(\Bmin,\,\Bmax)$, so $\gCS$ must vanish everywhere,
proving the theorem.

\section{Quasisymmetry in various coordinate systems}
\label{a:Hamada}
In this appendix, we prove that $B$ has a single helicity in Boozer
coordinates if and only if $B$ has a single helicity in Hamada coordinates.
While one proof can be found in Ref. \onlinecite{SugamaNishimura}, here we
prove a more general result, that symmetry is
equivalent for any straight-field-line coordinate system in which the Jacobian is
proportional to some power of $B$.

A number of identities must be demonstrated before the main theorem is proved.
Begin by considering two sets of
coordinates, $( {\theta_x ,\zeta_x })$ and $( {\theta_y ,\zeta_y } )$, which are not necessarily Boozer or Hamada coordinates.
If the the coordinates are straight-field-line coordinates, then
$\vect{B}=\nabla \psi  \times \nabla \theta_x +\iotab \nabla
\zeta_x \times \nabla \psi
 =\nabla \psi  \times \nabla \theta_y +\iotab \nabla \zeta_y
\times \nabla \psi$
where $2\pi \psi  $ is the toroidal flux. Suppose the transformation from one
system to the other is written as
$\theta_y =\theta_x +\F \left( {\psi  ,\theta_x ,\zeta_x } \right)$,
$\zeta_y =\zeta_x +\G \left( {\psi  ,\theta_x ,\zeta_x } \right)$
where $\F $ and $\G $ are periodic in both the poloidal and toroidal angles.
Then
\begin{equation}
\label{ah:eq4}
\nabla \psi  \times \nabla \F +\iotab \nabla \G \times \nabla \psi
=0.
\end{equation}
The $\nabla \theta_x $ component of this equation tells us $\partial
\F /\partial \zeta_x =\iotab \mbox{\thinspace }\partial \G /\partial
\zeta_x $, so upon integrating,
$\F =\iotab \G +y\left( {\psi  ,\theta_x } \right)$.
Here and throughout this appendix, $\partial /\partial \zeta_x $ holds
$\theta_x $ fixed, $\partial /\partial \theta_x $ holds $\zeta_x $ fixed,
$\partial /\partial \zeta_y $ holds $\theta_y $ fixed, and $\partial
/\partial \theta_y $ holds $\zeta_y $ fixed. The $\nabla \zeta_x $
component of (\ref{ah:eq4}) implies $\partial \F /\partial \theta_x =\iotab\mbox{\thinspace }\partial \G /\partial \theta_x $, so
$\F =\iotab \G +w\left( {\psi  ,\zeta_x } \right)$.
By comparing these two relations between $\F $ and $\G $, $\F $ must equal $\iotab \G $ plus a flux
function, so
\begin{equation}
\label{ah:eq9}
\theta_y =\theta_x +\iotab \G  +\mathcal{A}\left( \psi  \right)
\,\,\mbox{ and }\,\,\zeta_y =\zeta_x +\G
\end{equation}
for some flux function $\mathcal{A}\left( \psi  \right)$.

Now suppose the Jacobian for the $\left( {\theta_x ,\zeta_x } \right)$
coordinates is proportional to $B^x$, that is,
$\nabla \psi  \cdot \nabla \theta_x \times \nabla \zeta_x =A_x \left( \psi \right)B^x$
for some flux function $A_x \left( \psi  \right)$. Hamada coordinates have
$x=0$ and Boozer coordinates have $x=2$. The flux surface average of a
quantity $Q$ is
\begin{equation}
\label{ah:eq10}
\left\langle Q \right\rangle =\frac{1}{{V}'}\int_0^{2\pi } {d\theta_x }
\int_0^{2\pi } {d\zeta_x } \frac{Q}{\nabla \psi  \cdot \nabla \theta_x
\times \nabla \zeta_x }
\end{equation}
where $V\left( \psi  \right)$ is the volume enclosed by a flux surface, and
the prime denotes $d/d\psi  $. Observe that
$\left\langle {B^x} \right\rangle =4\pi^2/({V}'A_x )$,
so
\begin{equation}
\label{ah:eq12}
\nabla \psi  \cdot \nabla \theta_x \times \nabla \zeta_x =\frac{4\pi
^2}{{V}'}\frac{B^x}{\left\langle {B^x} \right\rangle }.
\end{equation}

Now suppose the two straight-field-line coordinate systems $( {\theta_x
,\zeta_x } )$ and $( {\theta_y ,\zeta_y } )$ have Jacobians proportional to $B^x$ and $B^y$ respectively.
Then from (\ref{ah:eq12}),
\begin{equation}
\label{ah:eq14}
\nabla \psi  \cdot \nabla \theta_y \times \nabla \zeta_y
=
\frac{\left\langle {B^x}\right\rangle }{B^x}
\frac{B^y}{\left\langle {B^y} \right\rangle }
\nabla \psi  \cdot \nabla \theta_x \times \nabla \zeta_x .
\end{equation}

Next, we apply the chain rule to $\G $ from (\ref{ah:eq9}):
\begin{equation}
\label{ah:eq15}
\frac{\partial \G }{\partial \theta_y }=\frac{\partial \theta_x }{\partial
\theta_y }\frac{\partial \G }{\partial \theta_x }+\frac{\partial \zeta_x
}{\partial \theta_y }\frac{\partial \G }{\partial \zeta_x }.
\end{equation}
By applying $\partial /\partial \theta_y $ to (\ref{ah:eq9}), we find
$1=\partial \theta_x /\partial \theta_y +\iotab \partial
\G /\partial \theta_y $ and $0=\partial \zeta_x /\partial \theta_y
+\partial \G /\partial \theta_y $, so (\ref{ah:eq15}) implies
\begin{equation}
\label{ah:eq16}
\frac{\partial \G }{\partial \theta_y }=\left( {1-\iotab\frac{\partial \G }{\partial \theta_y }} \right)\frac{\partial \G }{\partial
\theta_x }-\frac{\partial \G }{\partial \theta_y }\frac{\partial \G }{\partial
\zeta_x }.
\end{equation}
Rearranging,
\begin{equation}
\label{ah:eq17}
\left( {1 + \frac{\partial \G }{\partial \zeta_x } +\iotab \frac{\partial \G }{\partial \theta_x}} \right)\frac{\partial \G }{\partial
\theta_y }=\frac{\partial \G }{\partial \theta_x }.
\end{equation}

Now, apply $\vect{B}\cdot \nabla $ to (\ref{ah:eq9}) to obtain
$\nabla \psi  \times \nabla \theta_y \cdot \nabla \zeta_y =\nabla \psi
\times \nabla \theta_x \cdot \nabla \zeta_x +\vect{B}\cdot \nabla \G $.
Noting (\ref{ah:eq14}), then
\begin{equation}
\label{ah:eq19}
\frac{\left\langle {B^x} \right\rangle }{B^x}\frac{B^y}{\left\langle {B^y}
\right\rangle }=1+\frac{\partial \G }{\partial \zeta_x }+\iotab\frac{\partial \G }{\partial \theta_x }.
\end{equation}
Substituting this expression into (\ref{ah:eq17}) then gives
\begin{equation}
\label{ah:eq20}
\frac{\left\langle {B^x}\right\rangle }{B^x}
\frac{B^y}{\left\langle {B^y} \right\rangle }
\frac{\p \G }{\p \theta_y}
= \frac{\p \G }{\p \theta_x}
\end{equation}
Repeating the steps from (\ref{ah:eq15})-(\ref{ah:eq17}) but applying $\p / \p\theta_y$ instead of $\p / \p\zeta_y$,
we can similarly derive
\begin{equation}
\label{ah:eq21}
\frac{\left\langle {B^x}\right\rangle }{B^x}
\frac{B^y}{\left\langle {B^y} \right\rangle }
\frac{\p \G }{\p \zeta_y}
= \frac{\p \G }{\p \zeta_x}
\end{equation}

Next, applying $\partial /\partial \theta_x $ to (\ref{ah:eq19}),
\begin{equation}
\label{ah:eq22}
\left( {y-x} \right)\frac{\left\langle {B^x} \right\rangle }{\left\langle
{B^y} \right\rangle }B^{y-x-1}\frac{\partial B}{\partial \theta_x }=\left(
{\frac{\partial }{\partial \zeta_x }+\iotab \frac{\partial
}{\partial \theta_x }} \right)\frac{\partial \G }{\partial \theta_x }.
\end{equation}
Recalling that $\vect{B}\cdot \nabla =\nabla \psi  \times \nabla \theta
_x \cdot \nabla \zeta_x \left[ {\left( {\partial /\partial \zeta_x }
\right)+\iotab \left( {\partial /\partial \theta_x } \right)}
\right]$, then (\ref{ah:eq22}) is equivalent to
\begin{equation}
\label{ah:eq23}
\vect{B}\cdot \nabla \frac{\partial \G }{\partial \theta_x }=\frac{4\pi
^2}{{V}'}\frac{\left( {y-x} \right)}{\left\langle {B^y} \right\rangle
}B^{y-1}\frac{\partial B}{\partial \theta_x },
\end{equation}
where we have also applied (\ref{ah:eq12}). We could have applied $\partial /\partial
\zeta_x $ to (\ref{ah:eq19}) instead of $\partial /\partial \theta_x $, and so it is
also true that
\begin{equation}
\label{ah:eq24}
\vect{B}\cdot \nabla \frac{\partial \G }{\partial \zeta_x }=\frac{4\pi
^2}{{V}'}\frac{\left( {y-x} \right)}{\left\langle {B^y} \right\rangle
}B^{y-1}\frac{\partial B}{\partial \zeta_x }.
\end{equation}

We are finally prepared to begin the main proof.
Suppose $B$ has only a single helicity in the $\left( {\theta_x ,\zeta_x }
\right)$ coordinates: $B=B\left( {M\theta_x -N\zeta_x } \right)$ for some
integers $M$ and $N$, or equivalently,
$N\, \partial B/\partial \theta_x +M\, \partial B/\partial \zeta_x=0$.
Then from (\ref{ah:eq23})-(\ref{ah:eq24}),
$\vect{B}\cdot \nabla \left( N\, \partial \G /\partial \theta_x +M\, \partial \G /\partial \zeta_x \right)=0$.
It follows that
$N\, \partial \G /\partial \theta_x +M\, \partial \G /\partial \zeta_x =S\left( \psi  \right)$
for some flux function $S\left( \psi  \right)$. Integrating this result in $\theta
_x $ and $\zeta_x $ from 0 to $2\pi $ in both variables, we find $S$ must be zero.
Then applying (\ref{ah:eq20}) and (\ref{ah:eq21}),
\begin{equation}
\label{ah:eq28}
N\, \partial \G /\partial \theta_y +M\, \partial \G /\partial \zeta_y =0.
\end{equation}
Finally, we form
\begin{eqnarray}
\label{ah:eq29}
 N\frac{\partial B}{\partial \theta_y }+M\frac{\partial B}{\partial \zeta
_y }
&=&N\left( {\frac{\partial \theta_x }{\partial \theta_y }\frac{\partial
B}{\partial \theta_x }+\frac{\partial \zeta_x }{\partial \theta_y
}\frac{\partial B}{\partial \zeta_x }} \right)+M\left( {\frac{\partial
\theta_x }{\partial \zeta_y }\frac{\partial B}{\partial \theta_x
}+\frac{\partial \zeta_x }{\partial \zeta_y }\frac{\partial B}{\partial
\zeta_x }} \right) \\
& =&N\left( {\left[ {1-\iotab \frac{\partial \G }{\partial \theta_y
}} \right]\frac{\partial B}{\partial \theta_x }-\frac{\partial \G }{\partial
\theta_y }\frac{\partial B}{\partial \zeta_x }} \right)+M\left(
{-\iotab \frac{\partial \G }{\partial \zeta_y }\frac{\partial
B}{\partial \theta_x }+\left[ {1-\frac{\partial \G }{\partial \zeta_y }}
\right]\frac{\partial B}{\partial \zeta_x }} \right).
\nonumber
\end{eqnarray}
The first equality above is the chain rule, and to get the second line we
have used the $\partial /\partial \theta_y $ and $\partial /\partial \zeta
_y $ derivatives of (\ref{ah:eq9}). The last line of (\ref{ah:eq29})
vanishes due to (\ref{ah:eq28}), and so
\begin{equation}
\label{ah:eq30}
N\frac{\partial B}{\partial \theta_x }+M\frac{\partial B}{\partial \zeta_x
}=0\mbox{\thinspace \thinspace }\Rightarrow \mbox{\thinspace \thinspace
}N\frac{\partial B}{\partial \theta_y }+M\frac{\partial B}{\partial \zeta
_y }=0.
\end{equation}
The right equality in (\ref{ah:eq30}) also implies the left one, since $x$ and $y$ were
arbitrary in the proof. For the specific case of $x=0$ and $y=2$, then $B$
has a single helicity in Boozer coordinates if and only if $B$ has a single
helicity in Hamada coordinates.

\section{Current in a general stellarator}
\label{a:current}
Here we calculate several relations which are satisfied by the current in any MHD equilibrium
with nested toroidal flux surfaces.
We begin by noting that the perpendicular current is $\vect{j}_\bot
=c\left( {d\ptot /d\psi } \right)B^{-2}\vect{B}\times \nabla \psi $,
where $\ptot$ is the sum of the pressures of each species and a flux function.

Next, recall that the coefficient $I\left( {\psi } \right)$ in the
covariant Boozer representation (\ref{covariant}) equals $2/c$ times the toroidal current
inside a flux surface. Therefore $I\left( {\psi } \right)=\left( {2/c}
\right)\int {d^2\vect{a}\cdot \vect{j}} $, where the surface
integral is performed over a constant-$\zeta $ cross-section of the plasma,
a surface which covers the region from magnetic axis out to the flux surface
$\psi $. The area element is $d^2\vect{a}=d\psi \,  d\theta \left(
{\nabla \psi \cdot \nabla \theta \times \nabla \zeta } \right)^{-1}\nabla
\zeta $.  Using $\vect{j}=(j_{||}/B)\vect{B}+\vect{j}_\bot$, (\ref{contravariant}), and (\ref{covariant}), we obtain
\begin{equation}
\label{aj:eq4}
I
=\frac{2}{c}\int_0^{\psi } {d\psi' } \left(
-c I\frac{d\ptot}{d\psi }\int_0^{2\pi } {d\theta } \frac{1}{B^2}
+ \int_0^{2\pi } {d\theta } \frac{j_{||}
}{B}
 \right),
\end{equation}
where everything in parentheses is evaluated at $\psi' $ rather than $\psi$.
We next integrate (\ref{aj:eq4}) over all $\zeta $. Recall that the flux surface
average in Boozer coordinates can be written
\begin{equation}
\label{aj:eq5}
\left\langle X \right\rangle =\frac{\int_0^{2\pi } {d\theta } \int_0^{2\pi }
{d\zeta } \left( {X/B^2} \right)}{\int_0^{2\pi } {d\theta } \int_0^{2\pi }
{d\zeta } \left( {1/B^2} \right)},
\end{equation}
and observe that $\left\langle {B^2} \right\rangle =4\pi
^2/\int_0^{2\pi } {d\theta } \int_0^{2\pi } {d\zeta } \left( {1/B^2}
\right)$.  Then differentiating (\ref{aj:eq4}) in $\psi$,
we can write
\begin{equation}
\label{aj:eq6m}
\frac{dI}{d\psi} +
\frac{4\pi I}{\left\langle {B^2} \right\rangle }\frac{d\ptot}{d\psi }
=\frac{4\pi
}{c} \frac{\langle {j_{\vert \vert } B}
\rangle }{\left\langle {B^2} \right\rangle }.
\end{equation}
The boundary condition for this equation is $I=0$ at the magnetic axis.

An analogous ODE for $G(\psi)$ can be derived by repeating the above analysis with a constant-$\theta$ surface:
\begin{equation}
\label{aj:eq7}
\frac{dG}{d\psi} +
\frac{4\pi G}{\left\langle {B^2} \right\rangle }\frac{d\ptot}{d\psi }
=-\frac{4\pi
}{c} \frac{\iotab \langle {j_{\vert \vert } B}
\rangle }{\left\langle {B^2} \right\rangle }.
\end{equation}
The boundary condition for $G$ is that it must go to its vacuum value
at the plasma edge.

If another equation for $\langle {j_{\vert \vert } B} \rangle $ in
terms of $I$ and $G$ can be obtained from kinetic theory (as we have done for an omnigenous stellarator in (\ref{eqb1})),
then this equation can be used with (\ref{aj:eq6m}) and (\ref{aj:eq7}) to calculate a self-consistent
current profile.


%

\end{document}